\documentclass[pra,twocolumn,superscriptaddress,showpacs,amsmath,amstex,amssymb,citeautoscript]{revtex4-1}

\usepackage[T1]{fontenc}
\usepackage[utf8]{inputenc}
\usepackage{lipsum}
\usepackage{amsmath}
\usepackage{amssymb}
\usepackage{bbm}
\usepackage{braket}
\usepackage{xcolor}
\usepackage{soul}
\usepackage{pifont}
\newcommand{\cmark}{\ding{51}}%
\newcommand{\xmark}{\ding{55}}%
\usepackage[mathscr]{euscript}
\usepackage[shortlabels]{enumitem}
\usepackage[justification=justified]{subcaption}
\usepackage{graphicx}
\usepackage{lipsum}
\allowdisplaybreaks
\usepackage{float}
\usepackage{graphicx}
\usepackage{dsfont}
\usepackage{comment}
\usepackage[colorlinks=true]{hyperref}  
\hypersetup{
    bookmarks=true,         
    unicode=false,          
    pdftoolbar=true,        
    pdfmenubar=true,        
    pdffitwindow=false,     
    pdfstartview={FitH},    
    pdftitle={Nodal Pairing},    
    pdfauthor={Christos, Sachdev, and Scheurer},     
    pdfsubject={},   
    pdfcreator={},   
    pdfproducer={}, 
    pdfkeywords={} {} {}, 
    pdfnewwindow=true,      
    colorlinks=true,       
    linkcolor=blue, 
    citecolor=blue,        
    filecolor=magenta,      
    urlcolor=blue           
}

\newcommand{\equref}[1]{Eq.~(\ref{#1})}
\newcommand{\equsref}[2]{Eqs.~(\ref{#1}) and (\ref{#2})}
\newcommand{\secref}[1]{Sec.~\ref{#1}}
\newcommand{\figref}[1]{Fig.~\ref{#1}}
\newcommand{\refcite}[1]{Ref.~\onlinecite{#1}}

\newcommand{\tableref}[1]{Table~\ref{#1}}

\newcommand{\pdagger}{{\phantom{\dagger}}}
\newcommand{\diff}{\mathrm{d}}

\renewcommand{\approx}{\simeq}

\renewcommand{\vec}[1]{\boldsymbol{#1}}

\definecolor{wrongultramarine}{rgb}{1,0.5,0}

\begin{document}

\title{Nodal band-off-diagonal superconductivity in twisted graphene superlattices}

\author{Maine Christos}
\affiliation{Department of Physics, Harvard University, Cambridge MA 02138, USA}

\author{Subir Sachdev}
\affiliation{Department of Physics, Harvard University, Cambridge MA 02138, USA}

\author{Mathias S.~Scheurer}
\affiliation{Institute for Theoretical Physics, University of Innsbruck, Innsbruck A-6020, Austria}
\affiliation{Institute for Theoretical Physics III, University of Stuttgart, 70550 Stuttgart, Germany}

\begin{abstract} 
The superconducting state and mechanism are among the least understood phenomena in twisted graphene systems. Recent tunneling experiments indicate a transition between nodal and gapped pairing with electron filling, which is not naturally understood within current theory. We demonstrate that the coexistence of superconductivity and flavor polarization leads to pairing channels that are guaranteed by symmetry to be entirely band-off-diagonal, with a variety of consequences: most notably, the pairing invariant under all symmetries can have  Bogoliubov Fermi surfaces in the superconducting state with protected nodal lines, or may be fully gapped, depending on parameters, and the band-off-diagonal chiral $p$-wave state exhibits transitions between gapped and nodal regions upon varying the doping. We demonstrate that band-off-diagonal pairing can be the leading state when only phonons are considered, and is also uniquely favored by fluctuations of a time-reversal-symmetric intervalley coherent order motivated by recent experiments. Consequently, band-off-diagonal superconductivity allows for the reconciliation of several key experimental observations in graphene moiré systems.
\end{abstract}

\maketitle

\section{Introduction}
The fascinating physics \cite{ReviewAnrei,ReviewBalents} of correlated graphene moiré superlattices, such as twisted bilayer (TBG) and twisted trilayer graphene (TTG), has generated extensive efforts to uncover the mysteries of their phase diagrams. Much progress has been made towards understanding their normal-state physics, including the correlated insulating phases \cite{Cao_2018,Lu_2019,Sharpe_2019,Nuckolls_2020,https://doi.org/10.48550/arxiv.2303.00024,Kang_2019,Bultinck_2020,Soejima_2020,Xie_2021,Kwan_2021,2022PhRvX..12b1018C,Xie_2021HF,ledwith2021tb,Wagner_2022,wang2022kekule,kwan2023electronphonon} and the reset behavior \cite{Wong_2020,Zondiner_2020}; the latter, which is believed to be associated with the onset of flavor polarization, appears in the same density range of and can coexist with superconductivity \cite{Wong_2020,Zondiner_2020,Park_2021,Hao_2021,DiodeEffect,morissetteElectronSpinResonance2022,PauliLimit,PhysRevB.98.054515,OurClassification,Lake_2022,Christos_2020,Khalaf_2021,2022PhRvX..12b1018C,PhysRevB.105.224508,Scammell_2022,PhysRevLett.127.247703,PhysRevB.105.094506}. 
However, the form and symmetry of the superconducting order parameter and the pairing glue are still unknown, despite significant theoretical efforts \cite{PhysRevLett.121.257001,PhysRevLett.122.257002,PhysRevLett.127.247703,PhysRevB.103.235401,PhononsAndBandren,2022arXiv220202353Y,PhysRevB.104.L121116,Wang_2021,PhysRevLett.129.187001,PhysRevB.103.L041103,2022arXiv220202353Y,2022arXiv220900007H,Khalaf_2021,Christos_2020,Lake_2022,OurClassification,Cea_2021,Kozii_2022,AshvinFluctuations}.

Tunneling conductance measurements taken within the superconducting state reveal V-shaped density of states (DOS) \cite{Oh_2021,TunnelingPerge} which can become U-shaped at other electron concentrations \cite{TunnelingPerge}. Setting aside the possibility of thermal fluctuations as origin \cite{2023arXiv230101344P}, this is most naturally interpreted as a transition from nodal to fully gapped superconductivity. For a consistent microscopic theoretical understanding, this provides the following challenges: (i) electron-phonon coupling---a widely discussed \cite{PhysRevLett.121.257001,PhysRevLett.122.257002,PhysRevLett.127.247703,PhysRevB.103.235401,PhononsAndBandren,2022arXiv220202353Y,PhysRevB.104.L121116} pairing mechanism in TBG and TTG---will typically mediate an entirely attractive interaction in the Cooper channel, with leading pairing state that transforms trivially under all symmetries and is thus fully gapped \cite{PhysRevB.90.184512,PhysRevB.93.174509}. (ii) Even when the low-energy interactions favor an irreducible presentation (IR), e.g., $E$ of $C_3$, with nodal basis functions ($p$- or $d$-wave), the generically fully gapped chiral configuration wins over the nodal nematic one within mean-field. (iii) Even if we assume that the nodal state is energetically favored, e.g., due to significant corrections beyond mean-field \cite{PhysRevLett.30.1108,PhysRevB.99.144507,PhysRevB.106.094509,OurClassification}, one is still left to explain why there is a transition to another, fully gapped superconductor upon changing the filling. 

In this work, we show that the combination of flavor polarization and the representations of the symmetries in the flat bands of TBG and TTG allow for pairing channels that are completely off-diagonal in the flat bands and that such band-off-diagonal states can naturally reconcile all three key challenges (i-iii). More specifically, we find two distinct band-off-diagonal states: one of them transforms under the trivial representation $A$ of the system's point group $C_6$ (or one of $A_{1,2}$ of $D_6$ if we set the displacement field to zero) but can nonetheless have symmetry-protected nodal lines, akin to Bogoliubov Fermi surfaces discussed in \cite{Brydon_2018,Agterberg_2017}, see \figref{fig:ToyModel}(a-c) for an intuitive visual explanation. The surprising possibility of the existence of such Bogliubov Fermi surfaces without external magnetic field is unique to twisted graphene systems in that it follows as a direct consequence of both the symmetry and relative flatness of their normal-state bands. The second off-diagonal state transforms under a two-dimensional IR ($E_2$ of $C_6$). Its associated chiral state, $E_{2}(1,i)$, which is favored in mean-field over the nematic one, has the unique property of exhibiting nodal lines or being fully gapped depending on the filling fraction, even when the order parameter is kept fixed. We supplement our general symmetry arguments and phenomenological models with Hartree-Fock (HF) calculations on the continuum model, studying a variety of different pairing mechanisms. We find that nodal band-off-diagonal pairing is favored by the optical $A_1$ and $B_1$ phonon modes and by fluctuations of a time-reversal symmetric intervalley coherent (T-IVC) state (the T-IVC state has Kekul\'e order on the 
graphene scale \cite{ReadSachdev90,LeeSachdev15,YazdaniQuantumHall}). Evidence for the former has been provided by a recent photoemission study \cite{chen2023strong} while evidence for the latter comes from recent STM experiments \cite{https://doi.org/10.48550/arxiv.2303.00024}. Furthermore, also fluctuations of a time-reversal-symmetric sublattice polarized state (SLP$+$) are attractive in the band-off-diagonal channel (see \tableref{PreferredPairingStates} for a formal definition of the order parameters). We also show that fluctuations of both T-IVC and of a nematic, time-reversal symmetric IVC order \cite{OurNematicTheory} favor either the band-off-diagonal $A$ or an $E_1$ state with band-diagonal components, which may also be nodal; the winner is determined by the relative amount of nematic IVC and T-IVC fluctuations.

\section{Results}\label{GeneralConsiderations}
\subsection{Possible pairing states}
Let us begin by classifying the superconducting instabilities in graphene moir\'e systems in the limit where the low-energy bands are spin polarized but allowing for multiple bands.
We denote the spinless low-energy fermionic creation operators by $c^\dagger_{\vec{k},\alpha,\eta}$ with momentum $\vec{k}$ in valley $\eta=\pm$, and of band index $\alpha$ labeling the upper ($\alpha=+$) and lower ($\alpha=-$) quasi-flat bands. As a result of two-fold rotational symmetry, $C_{2z}$, along the out-of-plane ($z$) direction or effective spinless time-reversal symmetry, $\Theta$, the non-interacting band structure $\xi_{\vec{k},\alpha,\eta}$ obeys $\xi_{\vec{k},\alpha,\eta}=\xi_{-\vec{k},\alpha,-\eta}\equiv \xi_{\eta\cdot \vec{k},\alpha}$ and intervalley pairing is expected to dominate. A general pairing order parameter in the inter-valley channel couples as
\begin{equation}
    \mathcal{H}_{\text{p}} = \sum_{\vec{k},\eta=\pm,\alpha,\alpha'} c^\dagger_{\vec{k},\alpha,\eta} \left( \Delta_{\vec{k},\eta}\right)_{\alpha,\alpha'} c^\dagger_{-\vec{k},\alpha',-\eta} + \text{H.c.}, \label{CouplingOfElToSC}
\end{equation}
where the order parameter $\Delta_{\vec{k},\eta}=-\Delta^T_{-\vec{k},-\eta}$ is a matrix in band space. The physical spin texture of the superconductor is entirely determined by the form of the underlying normal state's polarization: if the spins are aligned in the two valleys, the superconductor is a non-unitary triplet, while anti-alignment \cite{morissetteElectronSpinResonance2022,Lake_2022} leads to a singlet-triplet admixed state \cite{OurClassification,Lake_2022,2022PhRvX..12b1018C}. In both cases, all of the following states are well defined, with the aforementioned spin structures and symmetries given by appropriate combinations of spinless operations and spin rotations (see Appendix A1).

We will classify the pairing states according to the irreducible representations (IRs) of the system's point group $D_{6}$, which is generated by six-fold rotations ($C_{6z}$) along the $z$ axis and two-fold rotation symmetry ($C_{2x}$) along the in-plane $x$ axis. Note a displacement field ($D_0 \neq 0$) breaks the in-plane rotations leading to the point group $C_6$. Importantly, all IRs of $D_{6}$ and $C_6$ are either even or odd under $C_{2z}$. Choosing the phases of the Bloch states such that $C_{2z}$ acts as
$c_{\vec{k},\alpha,\eta} \rightarrow c_{-\vec{k},\alpha,-\eta}$, it holds
\begin{equation}
    C_{2z}:\quad \Delta_{\vec{k},\eta} \quad  \longrightarrow \quad \Delta_{-\vec{k},-\eta} = -\Delta^T_{\vec{k},\eta}. \label{C2zSymOnPairing}
\end{equation}
This immediately implies that the pairing states in all IRs even under $C_{2z}$ ($A_1$, $A_{2}$, $E_2$ of $D_6$) must be anti-symmetric in band space and, thus, entirely band off-diagonal, whereas the order parameters of the other IRs ($B_1$, $B_{2}$, $E_1$) are symmetric and can contain both band-diagonal and band-off-diagonal components. 
While superconducting order parameters with finite band-off-diagonal components are rather common in multi-band systems, the existence of pairing states that are constrained to be entirely band-off-diagonal is rather unique and follows from the combination of $C_{2z}$ symmetry and the spin polarization in the normal state.
Importantly, this is unaffected by strain or nematic order breaking $C_{3z}$ as long as $C_{2z}$ remains, which guarantees that there are IRs with entirely band-off-diagonal order parameters.

\begin{table}[tb]
\begin{center}
\caption{Summary of pairing states in spin-polarized flat bands. Here $\chi_{\vec{k}}$ ($\hat{\chi}_{\vec{k}}$) is a real-valued (real and symmetric $2\times 2$ matrix-valued) MBZ-periodic function invariant under $C_{3z}$. Furthermore, $X_{\vec{k}}$ and $Y_{\vec{k}}$ ($\hat{X}_{\vec{k}}$ and $\hat{Y}_{\vec{k}}$) transform as $x$ and $y$ under $D_3$, generated by $C_{3z}$ and $C_{2x}$, while also being real (and symmetric). The third column indicates the type of nodes---line (ln), point (pt), or none (n)---on a generic Fermi surface for sufficiently small/large order-parameter magnitudes; options separated by ``or'' indicates that this depends on the normal-state band splitting, see main text. The last column shows which states merge when $D_0\neq 0$, reducing the point group from $D_6$ to $C_6$.}
\label{FormOfPairingStates}
\begin{ruledtabular}
 \begin{tabular} {ccccccc} 
IR of $D_6$ & $\Delta_{\vec{k},\eta} = -\Delta^T_{-\vec{k},-\eta}$ & nodes & IR of $C_6$ \\ \hline
$A_1$ & $\sigma_y \chi_{\eta\cdot\vec{k}}$, $\chi_{C_{2x}\vec{k}} = -\chi_{\vec{k}}$ & ln/pt or ln & $A$ \\
$A_2$ & $\sigma_y \chi_{\eta\cdot\vec{k}}$, $\chi_{C_{2x}\vec{k}} = \chi_{\vec{k}}$ & ln/n & $A$ \\
$E_{2}(1,0)$ & $\sigma_y Y_{\eta\cdot\vec{k}}$ & ln/ln or pt & $E_{2}(1,0)$  \\ 
$E_{2}(0,1)$ & $\sigma_y X_{\eta\cdot\vec{k}}$ & ln/ln or pt & $E_{2}(1,0)$  \\ 
$E_{2}(1,i)$ & $\sigma_y \left(X_{\eta\cdot\vec{k}} + i \, Y_{\eta\cdot\vec{k}} \right)$ & ln/ln or n & $E_{2}(1,i)$ \\ \hline
$B_1$ & $\eta \hat{\chi}_{\eta\cdot\vec{k}}$, $\sigma_z\hat{\chi}_{C_{2x}\vec{k}}\sigma_z = \hat{\chi}_{\vec{k}}$ & n & $B$  \\
$B_2$ &  $\eta \hat{\chi}_{\eta\cdot\vec{k}}$, $\sigma_z\hat{\chi}_{C_{2x}\vec{k}}\sigma_z = -\hat{\chi}_{\vec{k}}$ & pt & $B$  \\ 
$E_{1}(1,0)$ & $\eta \hat{X}_{\eta\cdot\vec{k}} $ & pt & $E_{1}(1,0)$ \\
$E_{1}(0,1)$ & $\eta \hat{Y}_{\eta\cdot\vec{k}} $ & pt & $E_{1}(1,0)$ \\
$E_{1}(1,i)$ & $\eta \left( \hat{X}_{\eta\cdot\vec{k}} + i\hat{Y}_{\eta\cdot\vec{k}} \right)$ & n & $E_{1}(1,i)$
 \end{tabular}
\end{ruledtabular}
\end{center}
\end{table}

Choosing the phase conventions of the Bloch states such that $C_{2x}$ and $C_{3z}$ act as $c_{\vec{k},\alpha,\eta}\rightarrow (\sigma_z)_{\alpha\alpha} c_{(k_x,-k_y),\alpha,\eta}$ and $c_{\vec{k},\alpha,\eta} \rightarrow c_{C_{3z}\vec{k},\alpha,\eta}$, respectively, the resulting candidate order parameters are summarized in \tableref{FormOfPairingStates}. Note that a momentum-independent representation of $C_{2x}$ must be $\sigma_z$ due to the bands' eigenvalues at the $\Gamma$-M line, which in turn are connected to the topological obstruction of the flat bands \cite{TopologicalObstruction}.  The reality (Hermiticity) constraint in \tableref{FormOfPairingStates} on $\chi$, $X$, and $Y$ ($\hat{\chi}$, $\hat{X}$, and $\hat{Y}$) comes from the residual spinless time-reversal symmetry $\Theta$ of the normal state \cite{SelectionRules,PhysRevB.48.3304}. The two two-dimensional IRs $E_{1,2}$ are each associated with three pairing states---two nematic phases $E_{1,2}(1,0)$, $E_{1,2}(0,1)$ and one chiral state $E_{1,2}(1,i)$.

\subsection{Spectral properties}\label{SpectralProperties}
We here have the rather unique situation that there are pairing channels, associated with the IRs $A_{1,2}$ and $E_{2}$, where the pairing is constrained by $C_{2z}$ to be entirely band off-diagonal. One immediate very unusual consequence is that the superconducting order parameter transforming under the trivial representation ($A_1$) has a symmetry-imposed line of zeros along the $\Gamma$-M line, and hence a nodal point in the spectrum. This is related to the topology-induced non-trivial representation of $C_{2x}$ in band space. We refer to \refcite{2022arXiv220202353Y} for the discussion of other topological nodal points for pairing in obstructed TBG bands.
As we will show next, band-off-diagonal pairing leads to additional unusual spectral properties with far reaching consequences for graphene moir\'e systems. To this end, consider the following effective Hamiltonian, $\mathcal{H}_{\sigma_y} = \sum_{\vec{k}} c^\dagger_{\vec{k},\alpha,\eta} c^\pdagger_{\vec{k},\alpha,\eta} \xi_{\eta\cdot\vec{k},\alpha} + \sum_{\vec{k}} [ \Delta_{\vec{k}}\, c^\dagger_{\vec{k},+} \sigma_y c^\dagger_{-\vec{k},-} + \text{H.c.} ]$, 
where the scalar function $\Delta_{\vec{k}}$ describes the form of pairing. We will here study two cases which are conventionally considered to be fully gapped, (i) a momentum-independent ``$s$-wave state'' ($A_2$ or $A$ pairing in \tableref{FormOfPairingStates}) where $\Delta_{\vec{k}} = \Delta_0$ and (ii) a ``chiral $p$-wave'' state, or more precisely an $E_{2}(1,i)$ state, where $\Delta_{\vec{k}} = \Delta_0 (X_{\vec{k}} + i Y_{\vec{k}})$ with $(X_{\vec{k}},Y_{\vec{k}})$ being smooth, MBZ-periodic functions transforming as $(x,y)$ under $C_{3z}$. Furthermore, we parameterize the dispersion, $\xi_{\eta\cdot \vec{k},\alpha}$, of the two flat bands ($\alpha=\pm$) in valley $\eta=\pm$ as $\xi_{\vec{k},\alpha} = \epsilon_{\vec{k}}-\mu + \alpha \,\delta_{\vec{k}}$, 
where $\epsilon_{\vec{k}}$ and $\delta_{\vec{k}}$ are $C_{3z}$ (and, for $D_0=0$, $C_{2x}$) symmetric functions.

\begin{figure}[tb]
    \centering
    \includegraphics[width=\linewidth]{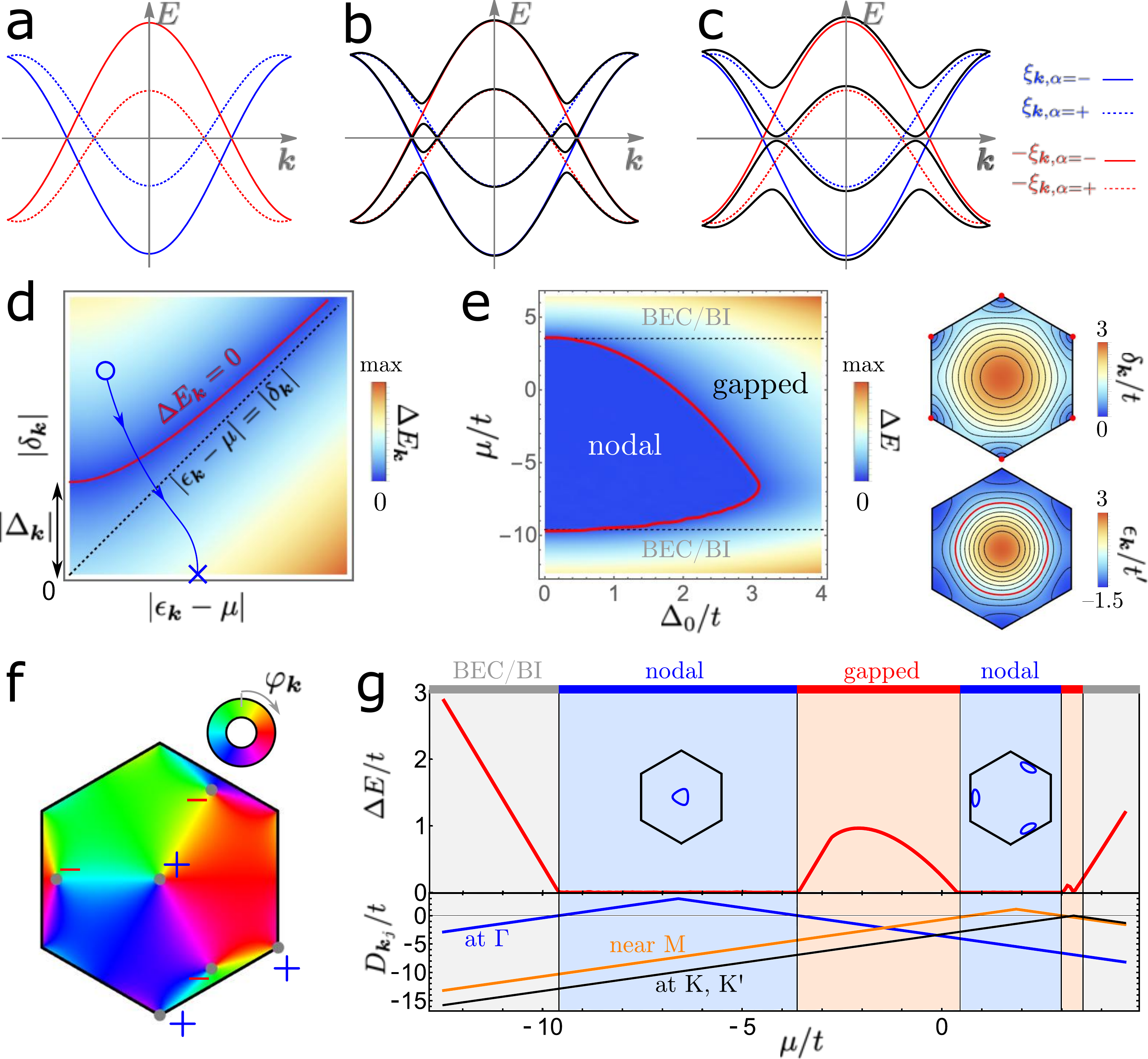}
    \caption{\textbf{Spectral properties of interband pairing.} While for band-diagonal pairing a small superconducting order parameter can immediately open up a gap as time-reversal symmetry guarantees that the associated avoided crossings [gray regions in (a)] in the Bogoliubov spectrum are at the Fermi level, this is not the case for band-off-diagonal pairing (b). Here, a sufficiently strong order-parameter value is required to establish a full gap, see (c). Its $\vec{k}$ dependence according to \equref{ExcitationSpectrum} is shown in (d), where the red line indicates nodal points. If the band structure has Dirac points, there will be a point on the horizontal axis (blue cross). Consequently, if there is another momentum point located above the red line (blue circle), continuity of the Hamiltonian implies a nodal point on any path connecting the two momenta. (e) Gap of the isotropic $A_2$ state and $\delta_{\vec{k}}$, $\epsilon_{\vec{k}}$ (zeros indicated in red) for the normal-state toy model defined in the text. BEC/BI refers to the Bose-Einstein condensate/band insulator limit. (f) Complex phase  $\varphi_{\vec{k}}=\text{arg}(X_{\vec{k}} + i Y_{\vec{k}})$ for leading basis function with small subleading corrections. (g) Shows the gap of the chiral $p$-wave $E_{2}(1,i)$ state with $\Delta_0=1.5t$ and the value of $D_{\vec{k}_j} := |\delta_{\vec{k}_j}|-|\epsilon_{\vec{k}_j}-\mu|$ for $\vec{k}_j$ at the three symmetry-in-equivalent vortices in (f)  as a function of $\mu$. We took $t'=-2.2t$, $t>0$, in (b,d).}
    \label{fig:ToyModel}
\end{figure}

The Bogoliubov spectrum of $\mathcal{H}_{\sigma_y}$ has four bands, given by $\pm \delta_{\vec{k}} \pm \sqrt{(\epsilon_{\vec{k}}-\mu)^2 + |\Delta_{\vec{k}}|^2}$. Consequently, the excitation gap at momentum $\vec{k}$ reads as
\begin{equation}
    \Delta E_{\vec{k}} = \left| |\delta_{\vec{k}}| - \sqrt{(\epsilon_{\vec{k}}-\mu)^2 + |\Delta_{\vec{k}}|^2} \right|, \label{ExcitationSpectrum}
\end{equation}
which is shown in \figref{fig:ToyModel}(d), and therefore exhibits nodes where $|\delta_{\vec{k}}|=\sqrt{(\epsilon_{\vec{k}}-\mu)^2 + |\Delta_{\vec{k}}|^2}$. As long as the band structure has Dirac points, there are points $\vec{k}_D$ in the Brillouin zone with $\delta_{\vec{k}_D}=0$, associated with the blue cross in \figref{fig:ToyModel}(d). Furthermore, for a metallic normal state, $\mu$ must be within the bandwidth and, hence, there must be a region $R$ in momentum space where $|\delta_{\vec{k}}| > |\epsilon_{\vec{k}}-\mu|$. For the momentum-independent $A_{2}$ state, $\Delta_{\vec{k}} = \Delta_0$, this implies that there exists $\Delta_0^c>0$ such that there is $\vec{k}^* \in R$ with parameters (such as the blue circle) above the red solid line in \figref{fig:ToyModel}(d) as long as $|\Delta_0| <\Delta_0^c$. By continuity, this means that there must be a nodal point on any line connecting $\vec{k}_D$ and $\vec{k}^*$. Consequently, for $\mu$ within the bandwidth and $\delta_{\vec{k}_D}=0$ for some $\vec{k}_D$, the $A_{2}$ will always have a nodal line if $|\Delta_0|$ is sufficiently small, consistent with the intuitive picture based on the Bogoliubov spectrum in \figref{fig:ToyModel}(a-c).

We illustrate this further in \figref{fig:ToyModel}(e) using a toy model with $\delta_{\vec{k}} = t \left| 1 + e^{i \vec{a}_1\cdot \vec{k}} + e^{-i \vec{a}_2\cdot \vec{k}}  \right|$ and $\epsilon_{\vec{k}} = t' \sum_{j=1}^3 \cos \vec{a}_j\cdot \vec{k}$, $\vec{a}_j = [C_{3z}]^{j-1}(\sqrt{3},0)^T$. 
This leads to the second unexpected conclusion that, for any pairing mechanism, including conventional electron-phonon coupling, the leading instability either has nodal lines in a finite region below $T_c$ or transforms non-trivially under the symmetries of the normal state. For electron-phonon pairing (or pairing mediated by the fluctuations of any time-reversal-symmetric order parameter \cite{PhysRevB.93.174509}, such as the T-IVC state) this is particularly unexpected since it is generally believed to always lead to a fully gapped state that transforms trivially under all symmetries. In fact, this can be proven in general terms \cite{PhysRevB.90.184512,PhysRevB.93.174509}, even for spin-orbit-split Fermi surfaces and beyond mean-field theory \cite{PhysRevB.93.174509}. The crucial difference to these works, however, is that spinfull time-reversal is broken in our case such that the Fermi-Dirac constraint is inconsistent with a non-sign-changing, band-diagonal pairing state. This leads to the unique situation that although electron-phonon coupling will lead to entirely attractive interactions in the Cooper channel, the superconducting energetics is frustrated: the dominant pairing state is determined by whether the energetic loss due to non-resonant band-off-diagonal Cooper pairs ($A_2$ pairing) or the costs from sign changes of the order parameter (such as $B_1$) are less harmful. We will demonstrate this explicitly by a model calculation in \secref{ElPhPairing} below, where either $A_2$ or $B_1$ is dominant, depending on the form of the electron phonon coupling.

Let us first, however, discuss the general spectral properties of the ``chiral $p$-wave'' state which is canonically expected to be fully gapped as long as the Fermi surfaces do not cross the zeros of $X_{\vec{k}} + i Y_{\vec{k}}$. Three of these zeros have to be at the $\Gamma$, $K$, and $K'$ points as a consequence of $C_{3z}$ symmetry. In the absence of fine-tuning, $X_{\vec{k}} + i Y_{\vec{k}}$ will have vortices at these points with vorticity $v=+1$. As can be seen in \figref{fig:ToyModel}(f), where we show the phase of $X_{\vec{k}} + i Y_{\vec{k}}$ using an admixture of the two lowest-order terms, the net vorticity of $+3$ at these high-symmetry points has to be compensated by anti-vortices at generic momenta. The lowest possible number is three $C_{3z}$-related vortices, which appear near the M points in \figref{fig:ToyModel}(f). If it holds $|\delta_{\vec{k}}| > |\epsilon_{\vec{k}}-\mu|$ at any of these zeros $\vec{k}=\vec{k}_j$, we obtain a point above the red line in \figref{fig:ToyModel}(d) and, thus, a nodal point along any contour between that $\vec{k}_j$ and $\vec{k}_D$; as opposed to the $A_2$ state, this holds irrespective of the value of $\Delta_0$ and therefore all the way to zero temperature. In summary, we find that also the $E_{2}(1,i)$ ``chiral $p$-wave'' state is not generically fully gapped but instead will exhibit a nodal line encircling any zero $\vec{k}_j$ of $X_{\vec{k}} + i Y_{\vec{k}}$ with $|\delta_{\vec{k}_j}| > |\epsilon_{\vec{k}_j}-\mu|$. This leads to an interesting filling dependence of the superconducting gap, as we illustrate in our toy model in \figref{fig:ToyModel}(g) along with the criterion $D_{\vec{k}_j}:=|\delta_{\vec{k}_j}| - |\epsilon_{\vec{k}_j}-\mu| >0$ evaluated at the vortices at $\Gamma$, K/K', and near M. Depending on $\mu$, $D_{\vec{k}}$ is positive only near the $\Gamma$ point or only in a region surrounding the vortices close to the M points, leading to nodal lines encircling $\Gamma$ and near the M points, respectively, as shown in the inset of \figref{fig:ToyModel}(g). These regimes are separated by a fully gapped region where $D_{\vec{k}} < 0$ for all $\vec{k}$, which could explain the fully gapped to nodal transition seen in tunneling experiments \cite{TunnelingPerge} when the filling fraction is changed. Note that $D_{\vec{k}_j} = - |\epsilon_{\vec{k}_j}-\mu| \leq 0$ for $\vec{k}_j$ at the K and K' points. In \figref{fig:ToyModel}(g), $D_{K} = D_{K'}$ vanishes close to the top of the band, which simply means that the Fermi surfaces cross the K, K' points and the superconductor has nodal points for this fine-tuned value of the chemical potential.

\subsection{Fluctuation-induced pairing}
Having discussed the unique energetics of pairing and spectral properties of the resulting superconductors in spin-polarized quasi-flat-bands with Dirac cones on a general level, we next study these aspects more explicitly by solving the superconducting self-consistency equations in the flat bands common to alternating-twist graphene systems. We will start with pairing induced by fluctuations of a nearby symmetry-broken phase. To this end, we will couple the low-energy electrons introduced in \equref{CouplingOfElToSC} to a collective bosonic field $\phi^\pdagger_j(\vec{q}) = \phi^\dagger_j(-\vec{q})$ via
\begin{equation}
    \mathcal{H}_{\phi} = \sum_{\vec{k},\vec{q},j} c^\dagger_{\vec{k}+\vec{q},\alpha,\eta} \lambda^j_{\alpha,\eta;\alpha',\eta'} c^\pdagger_{\vec{k},\alpha',\eta'} \phi_j(\vec{q}), \label{BosonFermiCoupling}
\end{equation}
where the Hermitian matrices $\lambda^j$ capture the nature of the correlated insulating phase; we here choose and normalize $\lambda^j$ such that $(\lambda^j)^2=\mathbbm{1}$. Both for twisted bi- \cite{Bultinck_2020} and trilayer graphene \cite{Christos_2020,Xie_2021HF,ledwith2021tb}, the stable phases emerging out of the $U(4)\times U(4)$ \cite{Bultinck_2020} manifold in the chiral-flat (decoupled) limit are natural candidates.
Integrating out the bosonic modes, we obtain an effective electronic interaction which in the for superconductivity relevant intervalley Cooper channel reads as
\begin{align}\begin{split}
    \mathcal{H}_{\text{int}}^\phi &= - \sum_{\vec{k},\vec{k}'} \chi_{\vec{k}-\vec{k}'}\mathcal{V}_{(\eta,\alpha,\beta),(\eta',\alpha',\beta')} \\ 
    &\quad\times c^\dagger_{-\vec{k},\beta,-\eta} c^\dagger_{\vec{k},\alpha,\eta} c^\pdagger_{\vec{k}',\alpha',\eta'} c^\pdagger_{-\vec{k}',\beta',-\eta'},  \label{InducedInteraction}
\end{split}  
\end{align}
with vertex
\begin{equation}
    \mathcal{V}_{(\eta,\alpha,\beta),(\eta',\alpha',\beta')} = t_{\phi}\sum_j [\lambda^{j}_{\beta,\eta;\beta',\eta'}]^* \lambda^{j}_{\alpha,\eta;\alpha',\eta'}, \label{InteractionVertex}
\end{equation}
$t_\phi = \pm 1$ encoding whether the order parameter is even or odd under time-reversal, $\Theta \phi_j(\vec{q}) \Theta^\dagger = t_\phi \phi_j(\vec{q})$, and $\chi_{\vec{q}}>0$ denoting the (static) susceptibility of $\phi_j$. 

Before discussing numerical results for the full model, we first focus on perfectly flat bands. In this limit, the leading superconducting instability within mean-field theory is given by the largest eigenvalue of $\mathcal{V}$ in \equref{InteractionVertex} viewed as a matrix in the multi-index $(\eta,\alpha,\beta)$. Furthermore, if there is an anti-symmetric, valley-off-diagonal matrix $D$ obeying (see Methods) 
\begin{equation}
    [D \eta_x,\lambda^j]_{-t_\phi}\equiv D\eta_x\lambda^j-t_\phi\lambda^j D \eta_x = 0, \label{CommutatorCrit}
\end{equation}
the associated leading superconducting order parameter in \equref{CouplingOfElToSC} is given by $(\Delta_{\vec{k},\eta})_{\alpha,\alpha'} = \delta_{\vec{k}} (D\eta_x)_{\alpha,\eta;\alpha'\eta}$ with $\delta_{\vec{k}}>0$; here $\eta_j$ denote Pauli matrices in valley space and the precise form of $\delta_{\vec{k}}$ is determined by $\chi(\vec{q})$.

\begin{figure*}[tb]
    \centering
    \includegraphics[width=\linewidth]{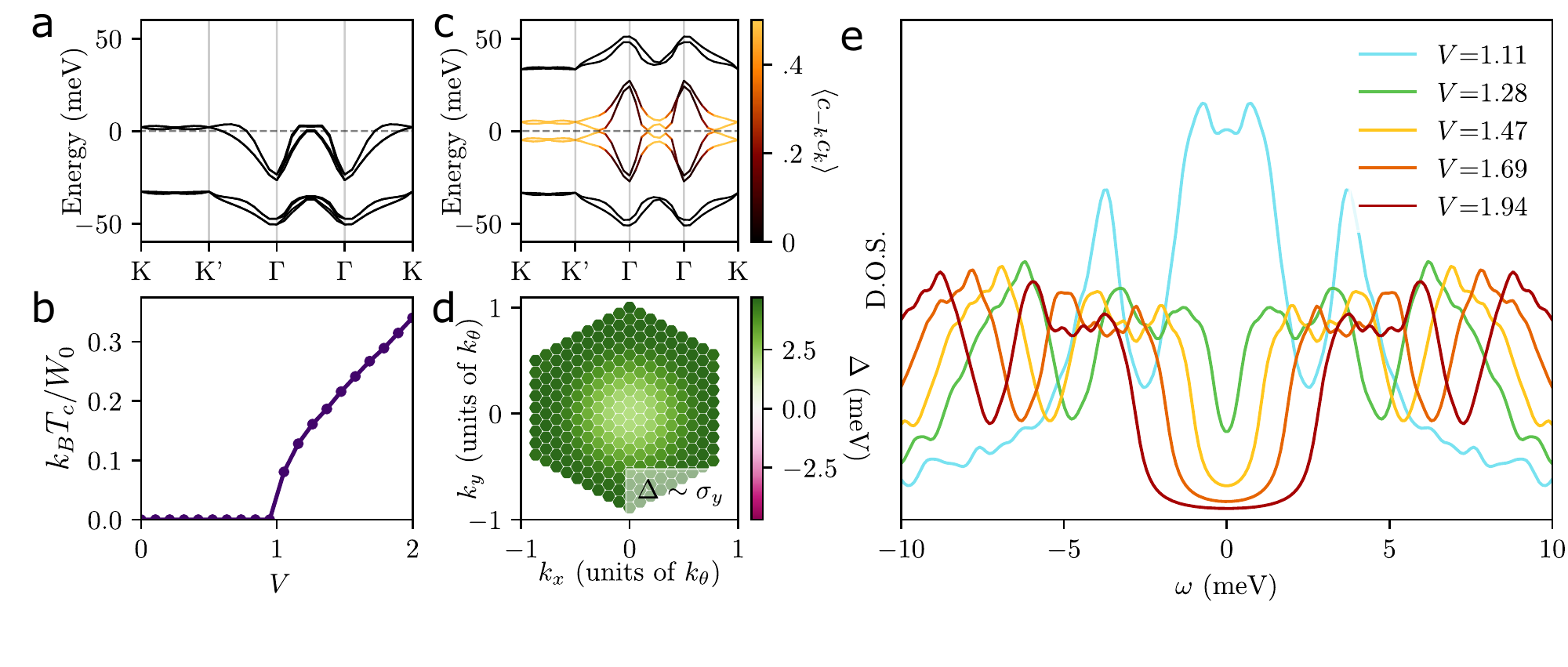}
    \caption{\textbf{Pairing mediated by T-IVC fluctuations.} We show (a) the band structure of the normal state with spin polarization (K, K', and $\Gamma$ label the high-symmetry points of the moir\'e scale Brillouin zone) and (b) the critical temperature $T_c$ (in units of the maximum band splitting $W_0\approx 9.4$ meV) as a function of coupling strength $V$ measured in units of the critical coupling $V_{c,1}=105\ \textrm{meV}\cdot\textrm{nm}^2$ obtained from the linearized gap equation. The band structure (with color indicating the band-projected value of the anomalous correlator) of the $A_2$ state and its order parameter are shown in (c) and (d). The DOS of the $T=0$ superconductor for several different values of coupling strength $V$ is plotted in (e). The DOS was computed as $\sum_{\vec{k}}\delta\left(E_{\vec{k}}-\omega\right)$, replacing the $\delta$ function with Lorentzians with half width at half max 0.3 meV (much smaller than the typical superconducting order parameter). The critical coupling $V_{c,2}$ where the nodal lines disappear is $V_{c,2}\approx1.4V_{c,1}$.}
    \label{fig:UnConventionalPairing}
\end{figure*}

\subsection{T-IVC fluctuations}\label{TIVC fluctuations}
Motivated by recent experiments \cite{https://doi.org/10.48550/arxiv.2303.00024} providing direct evidence for T-IVC order, we start with T-IVC fluctuations as a pairing glue. In the $U(4)\times U(4)$ symmetric limit, the T-IVC state is associated with $\lambda^j = \sigma_0 \eta_j$, $j=x,y$, within our conventions. Since $t_\phi=+1$, we are looking for $D \eta_x$ that commutes with $\lambda^j$. Interestingly, there is a unique anti-symmetric, valley-off-diagonal matrix $D\propto \sigma_y \eta_x$ with that property, implying that the leading pairing state has the form $\Delta_{\vec{k},\eta} = \sigma_y\delta_{\vec{k}}$, $\delta_{\vec{k}}>0$. This is exactly the $A_2$ state in \tableref{FormOfPairingStates}, which, as discussed above, will have nodal lines at least in the vicinity of $T_c$ when a finite band dispersion is taken into account. Intuitively, the fact that $A_2$ pairing is favored can be understood by noticing that the valley-off-diagonal form of $\lambda^j$ leads to an attractive interaction across the valleys, which penalizes the $B_1$ state with its sign change between the two valleys. In fact, it holds $\mathcal{V}_{(\eta,\alpha,\beta),(\eta',\alpha',\beta')} = (1-\eta\,\eta')\sum_{\mu=0}^3 (\sigma^*_\mu)_{\alpha,\beta} (\sigma_\mu)_{\alpha',\beta'}$ showing explicitly that it is repulsive (attractive) in the $B_1$ ($A_2$) channel.

To go beyond the flat-band limit, we solve the superconducting mean-field equations numerically. We take the flat TBG bands from the continuum model \cite{Bistritzer_2011} as the starting point. To capture the spin polarized normal state, we supplement it with Coulomb repulsion and a perform HF calculation (see Appendix A for details). As can be seen in the resulting band structure shown in \figref{fig:UnConventionalPairing}(a) with interaction renormalization assuming filling fraction $\nu=2$, this not only pushes one spin flavor below the Fermi level but also induces significant band renormalizations. For our subsequent study of superconducticity, we project onto the two bands at the Fermi level and associate them with the creation operators $c_{\vec{k},\alpha}$ in the interactions in \equsref{BosonFermiCoupling}{InducedInteraction}. 
In our numerical computations, we choose $\chi(\vec{q})=\frac{1}{A_{m}}\frac{V}{\alpha^2+|\vec{q}|^2/k_\theta^2}$
where $A_m$ is the real space area of a moir\'e unit cell, and take $\alpha=0.05$ for concreteness, although we checked our main conclusion do not crucially depend on this form. 
In all of our numerics, we work at doping $\nu=2.5$.

As expected, we indeed find that the $A_2$ state dominates, both right at the critical temperature $T_c$, obtained from the linearized gap equation, and at $T=0$ as we show by iteratively solving the full self-consistency equation (see Appendix C). One crucial effect of the finite dispersion and splitting between the bands is that a finite interaction strength, $V>V_{c,1}$, is required to stabilize the superconducting phase, as can be seen in the plot of $T_c$ in \figref{fig:UnConventionalPairing}(b). Superconductivity ceases to be a weak-coupling instability as the Bloch states $(\vec{k},\alpha,\eta)$ and $(-\vec{k},\alpha',-\eta)$ are not degenerate for $\alpha\neq \alpha'$, cutting off the logarithmic divergence known from BCS theory. The quasi-particle spectrum and order parameter of superconductivity from $T=0$ numerics are shown in Fig.~\ref{fig:UnConventionalPairing}(c,d). In accordance with our general discussion above, we observe that the order parameter only has finite components proportional to $\sigma_y$, which do not mix with the band-even contributions $\propto \sigma_{0,x,z}$ as a result of $C_{2z}$ symmetry. Furthermore, it does not change sign as a function of $\vec{k}$ and, for sufficiently small $V$ but still with $V>V_{c,1}$, the nodal lines in the superconducting spectrum persist all the way to $T=0$, while the nodal line is gapped out at low $T<T_c$ if $V>V_{c,2}$.

The interaction-strength-dependence of the superconducting gap can be more clearly seen in \figref{fig:UnConventionalPairing}(e), where we show the DOS for the self-consistent solution at $T=0$. For large $V$, the superconductor becomes fully gapped at $T=0$, leading to a U-shaped DOS. With smaller $V$, the magnitude of the order parameter decreases and the superconductor eventually exhibits nodal lines, as explained above. In the regime just before these nodal lines appear, there is an increase in the DOS near the Fermi level, roughly when the order parameter and the maximal band splitting are comparable, leading to a V-shaped DOS (green line). The lifetime parameter used to compute the DOS is $0.3$ meV; this choice was based on our $k$-grid spacing. While it is not necessarily small with respect to the tunneling gap (which vanishes at $V_{c,2}$), it is small with respect to $\Delta(\vec{k})$ which is of order 5 meV just as the state is becoming fully gapped for our choice of normal state. This behavior of the DOS with interaction strength may offer a natural explanation for the U-shaped tunneling conductance measurements near $\nu=2$ and V-shaped tunneling conductance measurements near $\nu=3$ observed in TTG \cite{TunnelingPerge}; if we are considering T-IVC fluctuations of the insulator at $\nu=2$, then it may be reasonable to expect the coupling to these fluctuations could grow weaker as we dope towards $\nu=3$, in line with the experimentally observed $\nu$ dependence. 

Note that the regime we call V-shaped here is strictly speaking fully gapped. However, the crucial difference to the BCS state is that the gap is much smaller than the order parameter magnitude as a result of the different Bogoliubov spectrum in \equref{ExcitationSpectrum}. This is why, depending not only on the magnitude of the pairing but also on the precise form of the normal state, the resulting tunneling spectra can resemble those observed experimentally \cite{Oh_2021,TunnelingPerge}, such as the green curve in \figref{fig:UnConventionalPairing}(e), making the $A_2$ state an attractive candidate. 
The regime of small $V$ where stable superconductivity with true Bogoliubov Fermi surfaces is observed can further exhibit a peak at $\omega=0$ which is due to a Van Hove singularity crossing the Fermi level, see blue curve in \figref{fig:UnConventionalPairing}(e); while this peak has not been observed experimentally, its presence crucially depends on details of the normal state band structure and is only found to be energetically favored in a very small regime of $V$ in our model.

\begin{figure*}[tb]
    \centering
    \includegraphics[width=\linewidth]{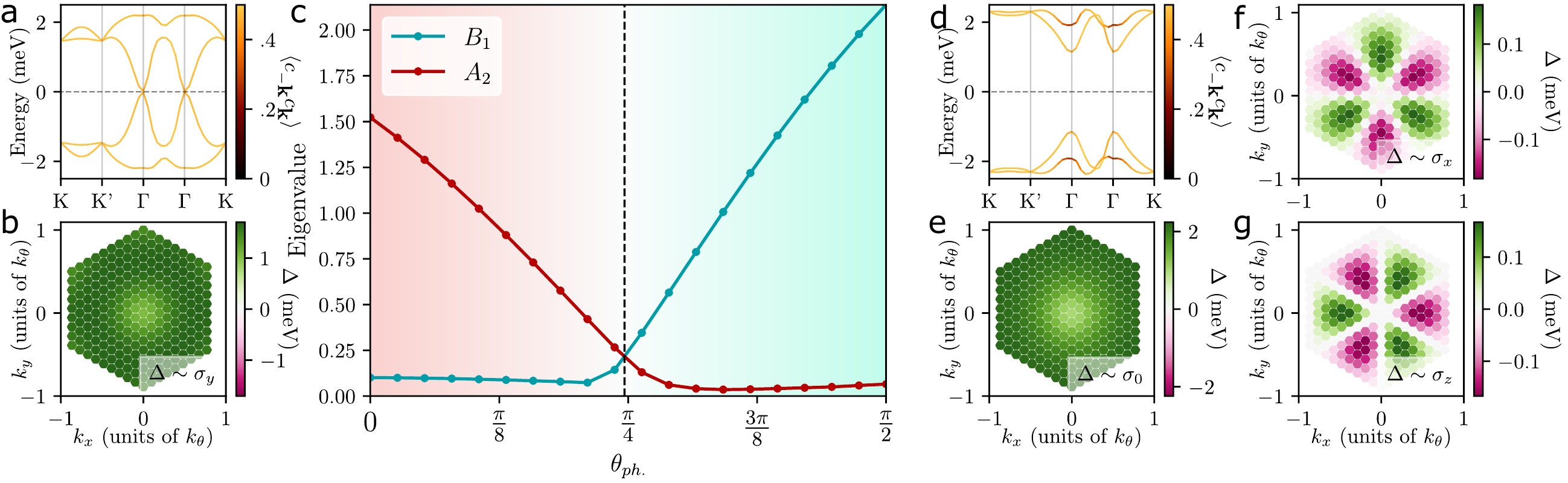}
    \caption{\textbf{Pairing from electron-phonon coupling.} 
    We show (a) the band structure and (b) the self consistent order parameter of the $A_2$ pairing for $\theta_{\text{ph}}=0$ and $T=0$. The eigenvalues corresponding to the $A_2$ and $B_1$ pairings in the linearized gap equation at $T= 5\,\textrm{K}$, which is close to their $T_c$, are shown in (c) as a function of $\theta_{\text{ph}}$. We show an example of the band structure (d) of the $B_1$ pairing and its order parameter (e,f,g). In accordance with symmetry, the $A_2$ ($B_1$) state only has order-parameter components $\propto \sigma_y$ ($\propto \sigma_{0,x,z}$). We took  $\nu=2.5$ and $V_0= 250$ meV $\cdot$ (nm)$^2$ with a continuum model bandwidth $\approx 2$ meV. We point out that if $A_1$ phonons are dominant, as suggested by recent experimental work \cite{chen2023strong} and past theoretical study in mono-layer graphene \cite{Basko_2008}, we would expect our $A_2$ pairing to dominate assuming the pairing potential is sufficiently large. We also emphasize that although the pairing functions for $A_2$ pairing (b) when $\theta=0$ and $B_1$ pairing when $\theta=\pi/2$ (e,f,g) are roughly equal, the excitation spectra shows the $B_1$ state with a band gap on the order of the pairing strength (d) while the $A_1$ state's band gap is nearly zero, see (a).}
    \label{fig:ConventionalPairing}
\end{figure*}

\subsection{Electron-phonon coupling}\label{ElPhPairing}
To illustrate that the off-diagonal $A_2$ state is more generally favored beyond just T-IVC fluctuations, we next discuss electron-phonon coupling, which is frequently considered as a plausible pairing mechanism for twisted moiré systems \cite{PhysRevLett.121.257001,PhysRevLett.122.257002,PhysRevLett.127.247703,PhysRevB.103.235401,PhononsAndBandren,2022arXiv220202353Y}. 
Similar to \refcite{PhysRevLett.121.257001}, we use that the optical $A_1$, $B_1$, and $E_2$ phonon modes are known \cite{PhysRevB.77.041409} to dominate the electron-phonon coupling in single-layer graphene. As these are optical phonons, we further assume that the impact of the interlayer coupling on the phonons can be neglected and arrive at
\begin{align}
    &\mathcal{H}_{EP} = \int \diff \vec{r} \, \psi^\dagger_{\ell,s}(\vec{r}) [ g_{A_1} \Lambda_{A_1} u_{A_1,\mu}(\vec{r}) \label{EPCHam} \\
    &+g_{B_1} \Lambda_{B_1} u_{B_1,\mu}(\vec{r}) + g_{E_2} \vec{\Lambda}_{E_2} \cdot \vec{u}_{E_2,\mu}(\vec{r})] (\vec{v}_\mu)_\ell\psi^\pdagger_{\ell,s}(\vec{r}) \nonumber
\end{align}
for the electron-phonon coupling, where $\vec{v}_\mu$ encode the layer structure of the modes (see Methods). Symmetry dictates that the vertices $\Lambda_g$ are given by $\Lambda_{A_1} = \eta_x \rho_x$, $\Lambda_{B_1} = \eta_y \rho_x$, and $\vec{\Lambda}_{E_2} = (\eta_z\rho_y,-\rho_x)$ where $\rho$ acts on the microscopic sublattice basis. Integrating out the phonons and projecting to the flat bands, we obtain an effective electron-electron interaction (see Methods)
\begin{align}\begin{split}
    \mathcal{H}^C_{\text{int}} &= -\sum_{\vec{k},\vec{k}'} V_g [\lambda^{g,j,\mu}_{\vec{k},\beta,\eta;\vec{k}',\beta',\eta'}]^* \lambda^{g,j,\mu}_{\vec{k},\alpha,\eta;\vec{k}',\alpha',\eta'} \\
    & \quad \times c^\dagger_{-\vec{k},\beta,-\eta} c^\dagger_{\vec{k},\alpha,\eta} c^\pdagger_{\vec{k}',\alpha',\eta'} c^\pdagger_{-\vec{k}',\beta',-\eta'}, \label{CooperChannelInteraction}\end{split}
\end{align}
where the coupling constants $V_g$ of the three different phonon modes $g=A_1,B_1,E_2$ are estimated to obey $V_{A_1} = V_{B_1} \approx 1.33 V_{E_2}$ for parallel spins in the two valleys, while $V_{A_1} = V_{B_1}=0$ for anti-parallel spins.
From \equref{CooperChannelInteraction}, it is clear that the induced interaction would be always completely attractive if we focused on intra-band pairing, $\alpha=\alpha'=\beta=\beta'$, which in spinful systems generically favors the trivial pairing channel \cite{PhysRevB.90.184512,PhysRevB.93.174509}. In our case, the combination of two energetically close bands and the trivial pairing being purely band-off-diagonal leads to the competition between different superconductors, even with electron-phonon coupling alone.

To demonstrate this, we study intra-valley pairing within the mean-field approximation and parametrize the relative strength of the different phonon modes with an angle variable $\theta_{\text{ph}}$ according to $V_{A_1}=V_{B_1} =  V_0 \cos \theta_{\text{ph}}$, $V_{E_2} =  V_0 \sin \theta_{\text{ph}}$. 
The results of the mean-field calculation are summarized in \figref{fig:ConventionalPairing}. We see that the $A_2$ pairing state is favored by the intervalley phonons ($\theta_{\text{ph}}=0$) inspite of its band-off-diagonal nature leading to a suppressed gap [see \figref{fig:ConventionalPairing}(a)]. This is natural as these phonons mediate an attractive interaction between the two valleys which disfavors the $B_1$ state, similar to T-IVC fluctuations. In fact, focusing on the leading, momentum independent term, $\lambda^{g,+}_{\vec{k},\vec{k}'}\rightarrow \lambda^{g,+}$, $g=A_1,B_1$, symmetry dictates $\lambda^{A_1,+} \propto \sigma_0 \eta_1$ and $\lambda^{B_1,+} \propto \sigma_0 \eta_2$ in the chiral limit (see Appendix D3). This maps the problem exactly to that of T-IVC fluctuations, immediately explaining why the order parameter has a fixed sign in \figref{fig:ConventionalPairing}(b). As $\theta_{\text{ph}}$ is increased, the $B_1$ state is favored (roughly for $\theta_{\text{ph}} > \pi/4$) as can be seen in \figref{fig:ConventionalPairing}(c).
This is expected since the intravalley $E_2$ phonon mediates an attractive interaction within each valley such that the energy gain due to the enhanced gap [\figref{fig:ConventionalPairing}(d)], associated with the band-diagonal matrix elements of the $B_1$ state, will overcompensate the energetic loss due to the sign change of $B_1$'s order parameter between the two valleys. This picture is consistent with the dominance and non-sign-changing nature of the band-diagonal components of the $B_1$ state, see \figref{fig:ConventionalPairing}(e-g). Finally, this behavior can also be understood by applying the commutator criterion in \equref{CommutatorCrit} in the microscopic sublattice basis, see Appendix D1.

This shows that, as opposed to the conventional scenario \cite{PhysRevB.90.184512,PhysRevB.93.174509}, there are two possible leading superconducting states and the superconducting pairing state does not transform trivially under the symmetries of the system even when phonons alone provide the pairing glue. We have checked in our $T=0$ numerics that a 60-70 meV$\cdot$(nm)$^2$ coupling to $A_1$ and $B_1$ phonons (based on \refcite{PhysRevB.77.041409}) is roughly of the order needed to stabilize the $A_2$ pairing, assuming the normal state is the flat bands of the un-renormalized continuum model, which in our case has a bandwidth of 2 meV. However, we note that if the interaction-renormalized band splitting is much larger than the continuum model band width, or if the normal state has anti-parallel spins in either valley, additional particle-hole fluctuations, such as those of T-IVC order, will also be required for pairing. An interesting scenario arises for anti-parallel spins in the two valley as a magnetic field will cant the spins and, hence, increase the projection of the intervalley phonon matrix elements to the flat bands. At least in TTG, with the suppressed orbital coupling, this could give rise to re-entrant superconductivity at high fields \cite{PauliLimit}.

\subsection{Other particle-hole fluctuations}

\begin{table}[tb]
\begin{center}
\caption{Leading superconducting states in the flat-band limit, following from \equref{CommutatorCrit}, for pairing mediated by fluctuations of the indicated orders, defined by using $\lambda^j$ in \equref{BosonFermiCoupling}. Here $\delta_{\vec{k}} > 0$ and states separated by commas are degenerate. The couplings in the microscopic basis, used in \figref{fig:MoreFluc} for the respective orders, are listed under $\bar{\lambda}^j$. Except for SLP$-$, the leading superconducting states for $\lambda^j$ and $\bar{\lambda}^j$ are the same (cf.~\figref{fig:MoreFluc} and Appendix D1).}
\label{PreferredPairingStates}
\begin{ruledtabular}
 \begin{tabular} {ccccc} 
\multicolumn{3}{c}{Fluctuating Order} & \multicolumn{2}{c}{Leading Superconductor} \\
\cline{1-3} \cline{4-5}
type & $\lambda^j$ & $\bar{\lambda}^j$ & $\Delta_{\vec{k},\eta}$ & IR
\\ \hline
T-IVC & $\sigma_0 \eta_{x,y}$ & $\rho_x \eta_{x,y}$ & $\sigma_y \delta_{\vec{k}}$ & $A_2$ \\
K-IVC & $\sigma_y \eta_{x,y}$ & $\rho_y \eta_{x,y}$ & $\sigma_0 \eta \delta_{\vec{k}}$ & $B_1$ \\
SLP$+$ & $\sigma_y \eta_{z}$ & $\rho_z \eta_0$ & $\sigma_y \delta_{\vec{k}}$, $\sigma_0 \eta \delta_{\vec{k}}$ & $A_2$, $B_1$ \\
SLP$-$ & $\sigma_y \eta_{0}$ & $\rho_z\eta_z$ & $\sigma_x \eta \delta_{\vec{k}}$, $\sigma_z \eta \delta_{\vec{k}}$ & $B_2$, $B_1$
 \end{tabular}
 \end{ruledtabular}
\end{center}
\end{table}

\begin{figure}[tb]
    \centering
    \includegraphics[width=\linewidth]{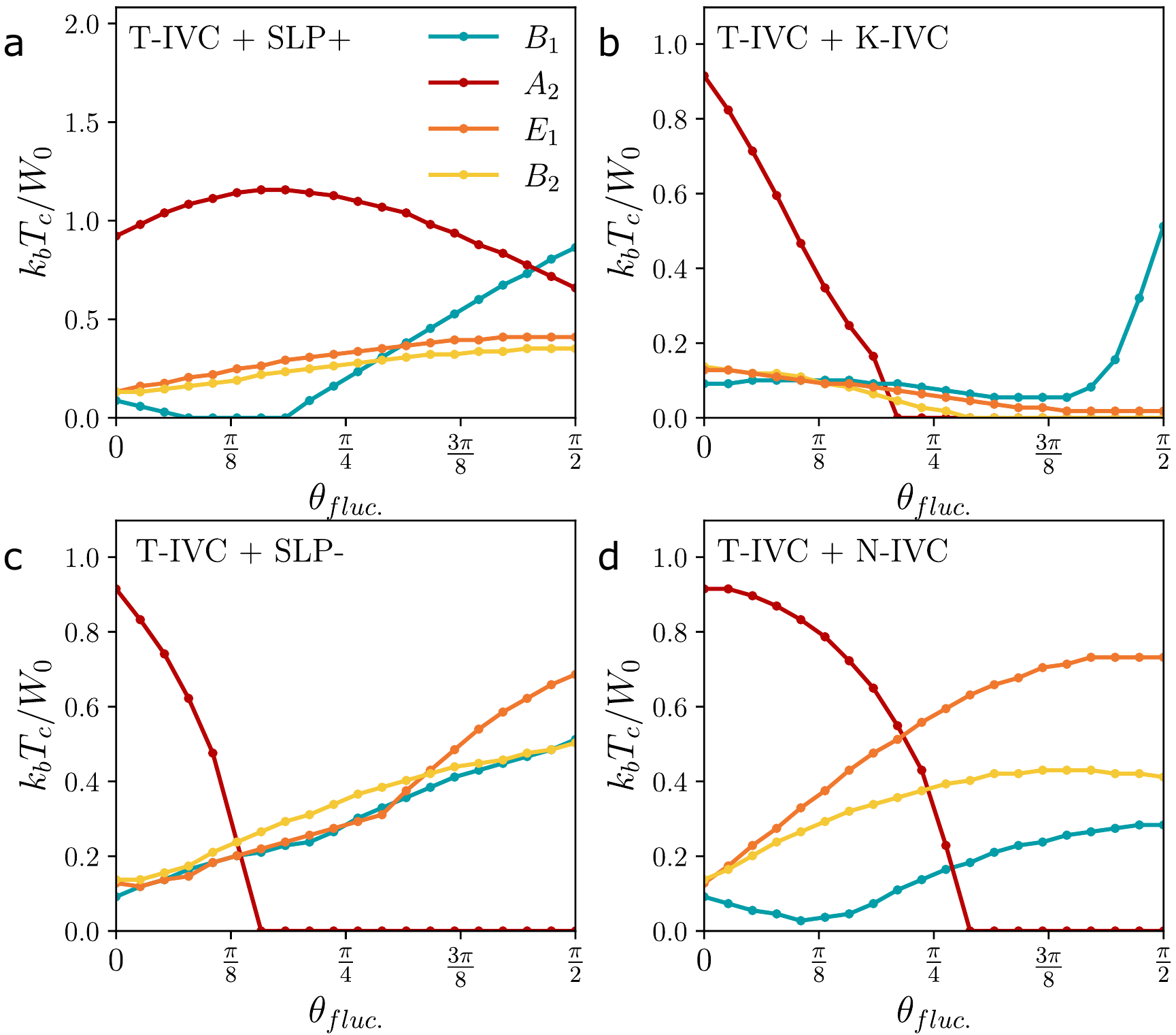}
    \caption{\textbf{Pairing for different particle-hole fluctuations.} These are defined by the coupling matrices $\bar{\lambda}^j$ listed in \tableref{PreferredPairingStates}. Similar to \figref{fig:ConventionalPairing}(c), we show $T_c$ of the leading pairing states, where $\theta_{\text{fluc.}}$ tunes the relative strength between T-IVC-induced interactions ($\propto \cos  \theta_{\text{fluc.}}$) and interactions ($\propto \sin  \theta_{\text{fluc.}}$) coming from fluctuations of (a) SLP$+$, (b) K-IVC, (c) SLP$-$, and (d) N-IVC fluctuations.}
    \label{fig:MoreFluc}
\end{figure}

Finally, we discuss pairing induced by fluctuations of other particle-hole instabilities. In \tableref{PreferredPairingStates}, we list the resulting leading superconductors taking $\lambda^j$ in \equref{BosonFermiCoupling} to be any of the different strong-coupling candidate order parameters \cite{Bultinck_2020,Christos_2020,2022PhRvX..12b1018C,Xie_2021HF,ledwith2021tb}. In particular, in addition to the T-IVC, we will consider the time-reversal-odd Kramers intervalley coherent state (K-IVC), and time reversal-odd and -even sublattice polarized states (SLP$-$ and SLP$+$). To analyze how sensitive our conclusions are to the precise form of the coupling of the  strong-coupling fluctuating orders to the electrons, we also perform numerics by projecting momentum-independent coupling vertices in the microscopic basis with the correct symmetries (see, e.g., Table II in \cite{2022PhRvX..12b1018C}),  listed as $\bar{\lambda}^j$ in \tableref{PreferredPairingStates}, to the flat bands. In the band basis, this leads to momentum-dependent coupling vertices, cf.~\equref{CooperChannelInteraction}. Motivated by recent experiments \cite{https://doi.org/10.48550/arxiv.2303.00024}, we will also consider fluctuations of an additional nematic, time-reversal symmetric, layer-odd, intervalley coherent state (N-IVC) \cite{OurNematicTheory} which is not a candidate ground state in the strong coupling limit; unlike the other strong-coupling ground states, the N-IVC has no momentum independent representation in the flat band basis but does have a momentum-independent matrix order parameter in the sublattice basis which takes the form $\lambda^{(j,j')} = (\eta_x,\eta_y)_j(\rho_0,\rho_z)_{j'}$. The results for fluctuations of the projected strong-coupling orders $\bar{\lambda}^j$ in \tableref{PreferredPairingStates} are shown in \figref{fig:MoreFluc}, where we use the angle $\theta_{\text{fluc.}}$ to tune the relative strength between T-IVC and any of the other type of fluctuation-induced interactions by multiplying the T-IVC interaction potential with $\cos(\theta_{\text{fluc.}})$ and the other fluctuation potential with $\sin(\theta_{\text{fluc.}})$. In our microscopic numerics, we have taken a potential form $\chi(\vec{q})=\frac{1}{A_{m}}\frac{V}{\alpha^2+|\vec{q}|^2/k_\theta^2}$ again with $\alpha=0.2$ and with $V=4200$ meV$\cdot$(nm)$^2$. We chose the value of $V$ such that the transitions between the different pairing states are clearly visible in \figref{fig:MoreFluc} when varying $\theta_{\text{fluc.}}$.  
In accordance with the prediction for $\bar{\lambda}^j$ in \tableref{PreferredPairingStates}, SLP$+$ fluctuations further stabilize the $A_2$ superconductor, see \figref{fig:MoreFluc}(a). As such, the band-diagonal $B_1$ superconducting channel, where SLP$+$ fluctuations are also attractive, can become the leading channel (favored over $A_2$ as a result of the finite bandwidth) only very close to $\theta_{\text{fluc.}}=\pi/2$. K-IVC fluctuations, however, are repulsive for $A_2$ pairing and favor the $B_1$ state more strongly.

So far, the strong-coupling ($\lambda^j$) and sublattice ($\bar{\lambda}^j$) form of the couplings in \tableref{PreferredPairingStates} lead to the same conclusions. This is different for SLP$-$ fluctuations [\figref{fig:MoreFluc}(d)], where the projection-induced momentum-dependence in the band basis can stabilize the $E_1$ superconductor. This can be understood by applying \equref{CommutatorCrit} in the sublattice basis (see Appendix D1). We also find the $E_1$ state when fluctuations of the N-IVC state of \refcite{OurNematicTheory} dominate. Examples of the $E_1$ nematic and $B_2$ order parameters which emerge for SLP$-$ fluctuations or N-IVC fluctuations are shown in Appendix F. We point out the nematic $E_1$ pairing is also an interesting candidate given that despite having nonzero pairing in the $\sigma_0$, $\sigma_x$, $\sigma_z$ channels, it will be nodal as long as the $\sigma_x$ components do not gap out the nodes in the band-diagonal parts.

\section{Discussion}
Taken together, we see that the proposed band-off-diagonal $A_2$ superconductor is an especially attractive candidate for TBG and TTG: first, it can lead to both V-shaped or U-shaped DOS, depending on lifetime parameters, the normal state, and the coupling strength $V$, see \figref{fig:UnConventionalPairing}(e). As these parameters might vary from sample to sample and within a sample (e.g., $V$ is expected to decrease upon doping further away from the insulator), this can naturally explain the tunneling data of \cite{Oh_2021,TunnelingPerge}. We emphasize however that at least at the level of our mean-field numerics, we only expect a V-shape in the regime where the superconducting pairing is of the order of the bandwidth; this is the regime, where although the pairing is finite and can be quite large, the gap in the superconducting spectrum is either just closing or very small relative to the pairing. Increasing the pairing further will lead to an evolution from $V$ to $U$ shaped while decreasing the pairing will eventually lead to a nodal Fermi surface and presumably a peak at zero energy in the DOS. Second, despite its interband nature, $A_2$ is the unique pairing state that is favored by fluctuations of two out of the four strong-coupling-candidates we consider for the correlated insulator, see \figref{fig:MoreFluc}(a-c). What is more, this includes the T-IVC state, signatures of which are observed in recent experiments \cite{https://doi.org/10.48550/arxiv.2303.00024}. Finally, it is also favored by the likely dominant \cite{chen2023strong,Basko_2008} optical intervalley phonon modes. We emphasize that, both in the case of fluctuating correlated insulators and phonons, the minimum attractive coupling needed to stabilize a purely band off diagonal state depends on the energy splitting between the two flat bands in the normal state; if the bands of our normal state are closer to degenerate, irrespective of the total bandwidth, the needed coupling to stabilize the $A_2$ pairing in mean-field will decrease. 

The other band-off-diagonal superconductor we identify transforms under the IR $E_2$, i.e., can be thought of as a $p$-wave state. Its spectral properties also agree well with experiment as the chiral configurations, $E_2(1,i)$, which is favored within mean-field theory over a nematic $E_2$ state, can also have nodal regions, depending on filling. As can be seen in \figref{fig:ToyModel}(g), this can lead to a transition from gapped to nodal when increasing the electron filling starting at $\nu\approx 2$. However, as opposed to the $A_2$ state, $E_2$ does not naturally appear as leading instability when considering optical phonons or fluctuations of any of the strong-coupling order parameters of the correlated insulator. While this makes it energetically less natural than $A_2$, we cannot exclude it since its phenomenology agrees well with experiment and since the precise form of the coupling of the dominant low-energy collective excitations are not known---significant momentum dependencies beyond $\lambda^j$ and $\bar{\lambda}^j$ in \tableref{PreferredPairingStates} could stabilize $E_2$ pairing as well. 
We also find in our numerics a nematic $E_1$ state which may be preferred over its chiral version in the presence of sufficient strain or due to fluctuation corrections \cite{PhysRevLett.30.1108,PhysRevB.99.144507,PhysRevB.106.094509,OurClassification}. We find the $E_1$ state is the leading instability of nematic IVC fluctuations and SLP$-$ fluctuations, and is a subleading instability of T-IVC fluctuations. The $E_1$ state is interesting in its own right, as it can also be nodal. 

As superconductivity might further coexist with T-IVC order \cite{https://doi.org/10.48550/arxiv.2303.00024}, we have checked (see Appendix E) that this does not alter our main observation: the preserved $C_{2z}$ symmetry still allows for entirely band-off-diagonal states, with transitions from nodal to full gapped, which are stabilized (among other fluctuations) by intervalley phonons.  

For the future, it will be interesting to go beyond mean-field and analyze the competition of our band-off-diagonal states with odd-frequency pairing, which we study in a follow-up work \cite{OurFollowUp}. It also seems promising to study Andreev reflection \cite{Oh_2021,TunnelingPerge} for our interband pairing scenario.
On a more general level, our work shows that the observation of nodal pairing in twisted graphene systems does not immediately exclude a chiral superconducting state nor an entirely electron-phonon-based pairing mechanism.
It illustrates that a microscopic understanding of the superconducting states in graphene moiré systems requires taking into account their intrinsically multi-band nature.

\vspace{1em}

\textit{Note added.} Just before posting our work, \refcite{2023arXiv230315551L} appeared online, which discusses pairing induced by $A_1$ phonons in spinful TBG bands.

\vspace{2em}

\begin{center}
    \textbf{Methods}
\end{center} 

\noindent \textbf{Flat-band limit.} To derive \equref{CommutatorCrit}, we take the flat-band limit, $\xi_{\vec{k},\pm} \rightarrow 0$, in the linearized gap equation. For the interaction defined in Eqs.~(\ref{BosonFermiCoupling}-\ref{InteractionVertex}), we get (with moiré cell area $A_m$)
\begin{align}\begin{split}
    (\Delta_{\vec{k},\eta})_{\beta,\beta'} &= t_{\phi} \frac{1}{4A_{m}T} \sum_{\vec{k}'} \chi_{\vec{k}-\vec{k}'} \\ &\times \sum_j [\lambda^{j}_{\beta',\eta;\alpha',\eta'}]^* \lambda^{j}_{\beta,\eta;\alpha,\eta'} (\Delta_{\vec{k}',\eta'})_{\alpha,\alpha'}. \label{FlatBandGapEquation}
\end{split}\end{align}
We define $(\hat{\Delta}_{\vec{k}})_{\alpha,\eta;\alpha',\eta'} := (\Delta_{\vec{k},\eta})_{\alpha,\alpha'} \delta_{\eta,\eta'}$ and note that finding the leading superconducting state according to \equref{FlatBandGapEquation} is equivalent to determining $\hat{\Delta}_{\vec{k}}$ that maximizes the functional
\begin{equation}
    \mathcal{F}[\hat{\Delta}_{\vec{k}}] := \frac{\sum_{\vec{k},\vec{k}',j}  \chi_{\vec{k}-\vec{k}'}\,t_{\phi}\text{tr}[\lambda^j\hat{\Delta}_{\vec{k}'}^\pdagger (\lambda^{j})^\dagger \hat{\Delta}_{\vec{k}}^\dagger]}{\sum_{\vec{k}} \text{tr}[\hat{\Delta}_{\vec{k}}^\dagger \hat{\Delta}_{\vec{k}}^\pdagger]}.
\end{equation}
Since $\chi_{\vec{k}-\vec{k}'}>0$, the maximum value will be reached if we can maximize $t_{\phi}\text{tr}[\lambda^j\hat{\Delta}_{\vec{k}'}^\pdagger (\lambda^{j})^\dagger \hat{\Delta}_{\vec{k}}^\dagger]$ for each $\vec{k}$, $\vec{k}'$, $j$ separately. As the Frobenius inner product $\braket{A,B}_F = \text{tr}[A^\dagger B]$ reaches its maximum (minimum) at fixed $\braket{A,A}$ and $\braket{B,B}$, if $A = c B$ with $c>0$ ($c<0$), $t_{\phi}\text{tr}[\lambda^j\hat{\Delta}_{\vec{k}'}^\pdagger (\lambda^{j})^\dagger \hat{\Delta}_{\vec{k}}^\dagger]$ is maximized if $\hat{\Delta}_{\vec{k}} = t_\phi c_{\vec{k},\vec{k}'} \lambda^j\hat{\Delta}_{\vec{k}'}^\pdagger (\lambda^{j})^\dagger$ with $c_{\vec{k},\vec{k}'} > 0$. For the ansatz $\hat{\Delta}_{\vec{k}} = \delta_{\vec{k}} D\eta_x$ (and \textit{assuming} for now that $\delta_{\vec{k}}$ has a fixed sign for all $\vec{k}$), this is obeyed if
\begin{equation}
    D\eta_x = t_\phi \lambda^{j} D\eta_x (\lambda^{j})^\dagger, \quad \forall j. \label{EquationForD}
\end{equation}
We state \equref{EquationForD} as the (anti)commutator condition (\ref{CommutatorCrit}) in the main text [equivalent if $(\lambda^j)^2 = \mathbbm{1}$], not only because it highlights the simple algebraic and basis independent nature of the condition but also since it emphasizes the similarities to the generalized Anderson theorem of \cite{Scheurer2016,PhysRevResearch.2.023140}.

If we can find a solution to \equref{EquationForD}, we know that the maximum (or at least one of the possibly degenerate maxima) of $\mathcal{F}[\hat{\Delta}_{\vec{k}}]$ is of the form of $\hat{\Delta}_{\vec{k}} = \delta_{\vec{k}} D\eta_x$ where $\delta_{\vec{k}}$ is obtained as the maximum of the reduced functional
\begin{equation}
    \widetilde{\mathcal{F}}[\delta_{\vec{k}}] := \frac{\sum_{\vec{k},\vec{k}'} \chi_{\vec{k}-\vec{k}'} \delta^*_{\vec{k}}\delta_{\vec{k}'}^{\phantom{*}} }{\sum_{\vec{k}} |\delta_{\vec{k}}|^2},
\end{equation}
or equivalently as the largest eigenvector of $\chi_{\vec{k}-\vec{k}'}$ viewed as a matrix in $\vec{k}$ and $\vec{k}'$. As $\chi_{\vec{k}-\vec{k}'} > 0$ (due to stability), the Perron-Frobenium theorem then immediately implies $\delta_{\vec{k}}>0$, in line with out assumption above and as stated in the main text.

\vspace{1em}

\noindent \textbf{Electron-phonon coupling.} To present more details on the electron-phonon coupling, the associated displacement operators in \equref{EPCHam} can be expressed in terms of canonical bosons, $b_{g,\alpha,\mu,\vec{q}}$, 
\begin{equation}
    (u_{g,\mu}(\vec{r}))_j = \sum_{\vec{q}} \frac{b_{g,j,\mu,\vec{q}} e^{i\vec{q}\cdot \vec{r}} + \text{H.c.}}{\sqrt{2N M \omega_g(\vec{q})}},
\end{equation}
where $j$ refers to the two components for the $E_2$ phonon (is idle for $A_1$, $B_1$), $M$ is the carbon mass, and $\omega_g(\vec{q})$ is the phonon dispersion, characterizing the phononic part of the Hamiltonian, $\mathcal{H}_{P} = \sum_{\vec{q}} \omega_g(\vec{q}) b^\dagger_{g,j,\mu,\vec{q}} b^\pdagger_{g,j,\mu,\vec{q}}$.

As for $(\vec{v}_\mu)_\ell$ in \equref{EPCHam}, $\ell=1,2$ refers to the physical graphene layer in the case of TBG. One can, in principle, choose any orthonormal basis; we will find it convenient to use the layer-exchange even and odd states, $\vec{v}_\pm = (1,\pm 1)^T/\sqrt{2}$. For TTG, the situation is more involved (see Appendix D2), but our arguments about which phonons are attractive in which pairing channels will hold for both systems.

We project $\mathcal{H}_{EP}$ in \equref{EPCHam} onto the two flat bands ($\alpha=\pm$) in each valley $\eta$ of the spin polarized continuum-model, leading to a coupling term similar to \equref{BosonFermiCoupling} with momentum-dependent coupling matrices, $\lambda^j \rightarrow \lambda^{g,j,\mu}_{\vec{k},\vec{k}'}$. Investigating the matrix elements $\lambda^{g,j,\mu}_{\vec{k},\vec{k}'}$, we notice that they almost vanish for the layer-odd intervalley ($A_1$, $B_1$) phonons, which can be understood as a consequence of chiral and particle-hole symmetry (see Appendix D3). The situation is the reverse for the intravalley ($E_2$) phonons, where the layer-even matrix elements are numerically small and the layer-odd matrix elements dominate. We therefore focus on layer-even (odd) intervalley (intravalley) phonon couplings.

Neglecting the momentum dependence in the phonon frequencies and retardation effects, the resulting electron-electron interaction in the inter-valley Cooper channel obtained by integrating out the phonons is given by \equref{CooperChannelInteraction}. Here, $V_g = g_g^2/(2N \omega_g^2) > 0$ and $V_{A_1} = V_{B_1} \approx 1.33 V_{E_2}$ results from $g_{A_1} = g_{B_1} \approx g_{E_2}$ and the  phonon frequencies estimated in \refcite{PhysRevB.77.041409}. Importantly, this only holds for parallel spins in the two valleys. For anti-parallel spins, the projection of the coupling matrices to the flat bands vanishes for the intervalley phonon modes $A_1$ and $B_1$ such that $V_{A_1} = V_{B_1}=0$.
\section*{Data Availability}
The data generated in this study are available in the Zenodo database under the accession code https://zenodo.org/record/8381555 and in the figshare repository https://doi.org/10.6084/m9.figshare.23897019.
\section*{Code Availability}
The codes used to generate the plots are available from the corresponding author on request.
\begin{acknowledgments}
M.S.S.~acknowledges funding by the European Union (ERC-2021-STG, Project 101040651---SuperCorr). Views and opinions expressed are however those of the authors only and do not necessarily reflect those of the European Union or the European Research Council Executive Agency. Neither the European Union nor the granting authority can be held responsible for them. M.C. and S.S. acknowledge funding by U.S. National Science
Foundation grant No. DMR-2002850. M.S.S.~thanks B.~Putzer for discussions. M.C. thanks P.~Ledwith and J.~Dong, and D.~Parker for helpful discussions.
\end{acknowledgments}

\section*{Author Contributions Statement}
M.C., S.S., and M.S.S. contributed to the research. M.C. and M.S.S. performed the numerical computations and wrote the paper.

\section*{Competing Interests Statement}
The authors declare no competing interests.

\onecolumngrid
\begin{appendix}
    \section{Normal-state}\label{sec:NormalState}

\subsection{Parallel and anti-parallel spins}\label{OrderParametersInSpinBasis}
We first discuss in more detail the spin structure of the superconducting states and the meaning of the symmetries of the effectively spinless bands, used in the main text to classify the superconducting states. We distinguish the two cases of (i) parallel spins in the two valleys and (ii) anti-parallel spins. 
To understand the physical meaning of the spinless symmetries of the main text, we start by listing the symmetries and their representations on the continuum-model operators $\psi_{\rho,\ell,\eta,s}(\vec{r})$ and band-operators $d_{\vec{k},\alpha,\eta,s}$ before normal-state polarization, where $\rho$, $\ell$, $\eta$, $s$, and $\alpha$ are indices for the sublattice, layer, valley, spin, and the two flat bands, while $\rho_j$, $\eta_j$, $s_j$, and $\sigma_j$ are Pauli matrices in sublattice, valley, spin, and band space, respectively:
\begin{enumerate}
    \item Two-fold rotation along $z$, $C_{2z}:$ $\psi(\vec{r})\rightarrow \eta_x\rho_x\psi(-\vec{r})$ and $d_{\vec{k}} \rightarrow \eta_x d_{-\vec{k}}$
    \item Spinless time-reversal, $\Theta$: $\psi(\vec{r})\rightarrow \eta_x\psi(\vec{r})$ and $d_{\vec{k}} \rightarrow \eta_x d_{-\vec{k}}$
    \item Spinful time-reversal, $\Theta_s$: $\psi(\vec{r})\rightarrow \eta_x i s_y \psi(\vec{r})$ and $d_{\vec{k}} \rightarrow \eta_x i s_y d_{-\vec{k}}$
    \item SO(3) spin-rotations, $R_s(\vec{\varphi})$: $\psi(\vec{r})\rightarrow e^{i\vec{\varphi}\cdot\vec{s}/2} \psi(\vec{r})$ and $d_{\vec{k}} \rightarrow e^{i\vec{\varphi}\cdot\vec{s}/2} d_{\vec{k}}$
    \item Global U(1) gauge symmetry, $U(\phi)$: $\psi(\vec{r})\rightarrow e^{i\phi} \psi(\vec{r})$ and $d_{\vec{k}} \rightarrow e^{i\phi}  d_{\vec{k}}$
    \item Three-fold rotation along $z$, $C_{3z}$: $\psi(\vec{r})\rightarrow e^{i\frac{2\pi}{3} \rho_z\eta_z} \psi(C_{3z}\vec{r})$ and $d_{\vec{k}} \rightarrow d_{C_{3z}\vec{k}}$
    \item Two-fold rotation along $x$, $C_{2x}$: $\psi(\vec{r})\rightarrow \rho_x \psi(C_{2x}\vec{r})$ and $d_{\vec{k}} \rightarrow \sigma_z d_{C_{2x}\vec{k}}$
\end{enumerate}
Except for $\Theta_s$ and $\Theta$, which are anti-linear, all representations are linear. In case (i) and assuming for concreteness that the active bands at the Fermi level of the flat bands are entirely spin-up ($s=\uparrow$), we simply define the fermionic operators of the main text as
\begin{equation}
    c_{\vec{k},\alpha,\eta} := d_{\vec{k},\alpha,\eta,\uparrow}.
\end{equation}
The remaining (non-trivial) symmetries then act as $C_{2z}: c_{\vec{k}} \rightarrow \eta_x c_{-\vec{k}}$, $\Theta: c_{\vec{k}} \rightarrow \eta_x c_{-\vec{k}}$, $U(\phi): c_{\vec{k}} \rightarrow e^{i\phi}  c_{\vec{k}}$, $C_{3z}: c_{\vec{k}} \rightarrow c_{C_{3z}\vec{k}}$, and $C_{2x}: c_{\vec{k}} \rightarrow \sigma_z c_{C_{2x}\vec{k}}$, exactly as in the main text.

The situation is more non-trivial in case (ii). Let us assume, for notational simplicity, that the spin polarization of the active flat bands in valley $\eta=+$ is $s=\uparrow$ and in valley $\eta=-$ is $\downarrow$. Accordingly, we define
\begin{equation}
    c_{\vec{k},\alpha,+} := d_{\vec{k},\alpha,+,\uparrow}, \quad c_{\vec{k},\alpha,-} := d_{\vec{k},\alpha,-,\downarrow}, \label{DefinitionOfcOpCaseii}
\end{equation}
as the effectively spinless fermionic operators used in the main text. It clearly holds, exactly as before, $U(\phi): c_{\vec{k}} \rightarrow e^{i\phi}  c_{\vec{k}}$, $C_{3z}: c_{\vec{k}} \rightarrow c_{C_{3z}\vec{k}}$, and $C_{2x}: c_{\vec{k}} \rightarrow \sigma_z c_{C_{2x}\vec{k}}$. However, $\Theta$ and $C_{2z}$ are explicitly broken and, thus, have to be replaced by appropriate combinations with other symmetries. Let us define
\begin{equation}
    \widetilde{\Theta} := U(-\pi/2) \Theta_s R_s(\pi \hat{\vec{e}}_z), \quad \widetilde{C}_{2z} := U(-\pi/2) C_{2z} R_s(\pi\hat{\vec{e}}_x), \label{DefinitionOfThetaAndC2zTilde}
\end{equation}
which are symmetries of the system and obey the same algebraic relations as the symmetries in the main text,
\begin{equation}
    \widetilde{\Theta}^2 = \widetilde{C}_{2z}^2 = \mathbbm{1}, \quad [\widetilde{\Theta},\widetilde{C}_{2z}] = 0, \quad [\widetilde{\Theta},C_{2x}] = [\widetilde{\Theta},C_{3z}] = [\widetilde{C}_{2z},C_{2x}] = [\widetilde{C}_{2z},C_{3z}] = 0.
\end{equation}
In fact, their representation on the fermions defined in \equref{DefinitionOfcOpCaseii} is exactly the same as that of $\Theta$ and $C_{2z}$ in the main text, $\widetilde{C}_{2z}: c_{\vec{k}} \rightarrow \eta_x c_{-\vec{k}}$ and $\widetilde{\Theta}: c_{\vec{k}} \rightarrow \eta_x c_{-\vec{k}}$. As such, for case (ii), the time-reversal symmetry $\Theta$ and two-fold-rotational symmetry $C_{2z}$ in the main text can be identified with $\widetilde{\Theta}$ and $\widetilde{C}_{2z}$ in \equref{DefinitionOfThetaAndC2zTilde}.
To illustrate this further and also explicitly discuss the spin structure of the order parameter, we transform the superconducting order parameter back to the $d$-fermions via \equref{DefinitionOfcOpCaseii},
\begin{equation}
    \mathcal{H}_{\text{p}}=\sum_{\vec{k}} c^\dagger_{\vec{k},\alpha,+} (\Delta_{\vec{k}})_{\alpha,\alpha'} c^\dagger_{-\vec{k},\alpha',-} = \frac{1}{2} \sum_{\vec{k}} d^\dagger_{\vec{k},\alpha,+,s} [(s_0+s_z)i s_y]_{s,s'} (\Delta_{\vec{k}})_{\alpha,\alpha'}  d^\dagger_{-\vec{k},\alpha',-,s'}, \label{ExplicitFormOfPairingI}
\end{equation}
which shows that we obtain an admixture of singlet and (unitary) triplet pairing. To demonstrate the action of $\widetilde{\Theta}$ and $\widetilde{C}_{2z}$ more explicitly and provide a consistency check, let us focus on $\Delta_{\vec{k}} = 2\Delta \sigma_y$, where \equref{ExplicitFormOfPairingI} becomes
\begin{equation}
    \mathcal{H}_{\text{p}} = \frac{\Delta}{2} \sum_{\vec{k}} d^\dagger_{\vec{k}} (i s_0\eta_y+s_z \eta_x)i s_y\sigma_y  d^\dagger_{-\vec{k}}. \label{SpecificExampleI}
\end{equation}
From \equref{DefinitionOfThetaAndC2zTilde}, we find the representations $\widetilde{C}_{2z}: d_{\vec{k}} \rightarrow \eta_x s_x d_{-\vec{k}}$ and $\widetilde{\Theta}: d_{\vec{k}} \rightarrow \eta_x s_x d_{-\vec{k}}$; applying this in \equref{SpecificExampleI}, we find that 
\begin{equation}
    \widetilde{C}_{2z}: \Delta \rightarrow \Delta, \quad \widetilde{\Theta}: \Delta \rightarrow -\Delta^*,
\end{equation}
exactly as in the main text.

For case (i), \equref{ExplicitFormOfPairingI} instead becomes
\begin{equation}
    \mathcal{H}_{\text{p}}=\sum_{\vec{k}} c^\dagger_{\vec{k},\alpha,+} (\Delta_{\vec{k}})_{\alpha,\alpha'} c^\dagger_{-\vec{k},\alpha',-} = \frac{1}{2} \sum_{\vec{k}} d^\dagger_{\vec{k},\alpha,+,s} [(s_x+is_y)i s_y]_{s,s'} (\Delta_{\vec{k}})_{\alpha,\alpha'}  d^\dagger_{-\vec{k},\alpha',-,s'},
\end{equation}
i.e., a non-unitary triplet state---as expected \cite{OurClassification} since this is the ``Hund's partner'' of the singlet-triplet admixed state in \equref{ExplicitFormOfPairingI}, obtained by an independent spin-rotation in the two valleys [SU(2)$_- \times$ SU(2)$_+$]. For $\Delta_{\vec{k}} = 2\Delta \sigma_y$ this yields
\begin{equation}
    \mathcal{H}_{\text{p}} = \frac{\Delta}{2} \sum_{\vec{k}} d^\dagger_{\vec{k}} (s_x \eta_x + i s_y \eta_x)i s_y\sigma_y  d^\dagger_{-\vec{k}}. 
\end{equation}
Again in accordance with the spinless formulation of the main text, we get $C_{2z}: \Delta \rightarrow \Delta$ and $\Theta: \Delta \rightarrow -\Delta^*$.

We finally note that the normal-state polarization also determines the spin-structure of the fluctuating orders in Table II and \tableref{ExtendedPreferredPairingStates}: switching between the two scenarios (i) and (ii) requires replacing an order parameter for the correlated insulator by its ``Hund's partner'' (see, e.g., Table II in \cite{2022PhRvX..12b1018C} for a complete list). As the system is believed to be close to the SU(2)$_- \times$ SU(2)$_+$ symmetric limit (the intervalley Hund's coupling was estimated to be smaller than $0.1\,\textrm{meV}$ in \cite{morissetteElectronSpinResonance2022}), the strength of fluctuations of Hund's partners is expected to be roughly the same. As such, both scenarios (i) and (ii) are consistent with a mechanism based on fluctuations of an order parameter of a correlated insulator. As mentioned in the main text, this is different for phonons, where only scenario (i) allows for intervalley phonons providing the pairing glue.

\subsection{Hartree-Fock numerics}
To capture the non-interacting band structure, we use a continuum-model description \cite{Bistritzer_2011},
\begin{equation}
     \mathcal{H}_0 = \int \diff \vec{r} \, \psi^\dagger_{\rho,\ell,\eta,s}(\vec{r}) \left[ h_\eta(\vec{\nabla},\vec{r}) \right]_{\rho,\ell;\rho',\ell'} \psi^\pdagger_{\rho',\ell',\eta,s}(\vec{r}),  \label{ContinuumModel} 
\end{equation}

where $\psi^\dagger_{\rho,\ell,\eta,s}$ creates an electron of spin $s=\uparrow,\downarrow$, in valley $\eta=\pm$, sublattice $\rho=A,B$, and with pseudo-layer quantum-number $\ell=1,2$; in case of TBG, $\ell$ refers to the actual two graphene layers, whereas, for TTG, it denotes the two mirror-even layer-eigenstates, $(1,1,1)^T$ and $(1,-2,1)^T$, of the three layers \cite{Khalaf_2019}. The continuum model involves two terms, $(h_\eta)_{\ell,\ell'}=\delta_{\ell,\ell'}h^{(d)}_{\ell,\eta}(\vec{\nabla}) + (h^{(t)}_{\eta}(\vec{r}))_{\ell,\ell'}$; the first one, $h^{(d)}_{\ell,\eta} = -i\hbar v_F e^{i \frac{\rho_z \theta_\ell}{2}} (\eta \rho_x \partial_x - \rho_y \partial_y) e^{-i \frac{\rho_z \theta_\ell}{2}}$ with $\rho_j$ being Pauli matrices in sublattice space, describes the Dirac cones of chirality $\eta$, rotated by $\theta_\ell=(-1)^\ell \theta/2$ in the two (pseudo)layers $\ell$; the second one, $h^{(t)}$, captures the tunneling between the layers, with amplitude $w_0$ and $w_1$ between the same and opposite sublattices, respectively. The modulation of the tunneling on the moir\'e scale leads to a reconstruction of the band structure, exhibiting nearly flat bands for magic angles around $\theta\approx 1.1^\circ$ and $\theta\approx 1.5^\circ$ for TBG and TTG, respectively. We take $w_1=89$ meV, $\frac{w_0}{w_1}=.55$, $v_F=10^6$ m/s, $\theta=1.09^{\circ}$ in all our numerical calculations.

As already mentioned above, experiments \cite{Zondiner_2020,Wong_2020} indicate that the superconducting phase in the density regime $2 < |\nu| < 3$ coexists with the reset behavior at half-filling, $|\nu | = 2$, of the upper or low flat-bands. To model this effect, we add Coulomb repulsion,
\begin{equation}
    \mathcal{H}_C = \frac{1}{2N}\sum_{\vec{q}} V(\vec{q}) \rho_{\vec{q}}\rho_{-\vec{q}} \label{CoulombInteraction}
\end{equation}
to our Hamiltonian, where $\rho_{\vec{q}}$ is the Fourier transform of the density of the continuum-model electrons $c_{\vec{r}}$ and the $N$ the number of moir\'e unit cells. We assume a double gate screened Coulomb potential of the form:
\begin{equation}
    V(\vec{q})=\frac{1}{A_m}\frac{1-e^{-2d_s|\vec{q}|}}{2\epsilon\epsilon_0|\vec{q}|}
\end{equation}
In the above, $A_m$ is the area of a real-space moir\'e unit cell (since we consider TBG and not TTG in our numerics, we take $A_m$ to be the moir\'e unit cell for 1.09$^\circ$), $d_s$ is the screening distance which we take to be 40 nm, and $\epsilon$ is the dielectric constant we take to be $\epsilon\approx 4$.
Note that projecting \equref{CoulombInteraction} into the bands of TTG will also lead to interactions coupling the mirror-sectors. However, as was shown \cite{2022PhRvX..12b1018C} analytically in a specific limit and numerically for realistic parameters, also the interacting physics of TTG decays into that of the TBG and that of a single Dirac cone for $D_0=0$. As such, it is justified to focus on the mirror-bands as in \equref{ContinuumModel} when discussing the reset physics in TTG at $D_0=0$.

In computing the normal state, we assume the same normal state density matrix as in \refcite{2022PhRvX..12b1018C} where the expectation value $\langle c^\dagger_{\vec{k},\alpha,\eta} c_{\vec{k},\beta,\eta}\rangle$ is equal to the $\frac{1}{2}$Id in the subspace of the flat bands of one spin flavor which are half filled in our normal state and equal to  Id in the flat bands of the remaining spin flavor which are fully polarized. We emphasize that we are assuming a static, momentum independent ansatz for the normal state density matrix which is not obtained self consistently.
As can be seen in Fig. II, instead of just rigidly shifting one spin species away from the Fermi level, there are also significant band renormalizations, in particular for the active spin flavor. Similar to the toy model with $t'<0$ used in Fig.~1, the Dirac cones at the K and K' points are pushed towards the top of the bands.
\subsection{Gauge Fixing}
We will also describe how we fix the phases of the continuum model Bloch wavefunctions we use in our computations. We denote the wavefunction of band $n$ in valley $\eta$ at momentum $\vec{k}$ by $u_{\vec{k},n,\eta}$. We use $C_{2z}\mathcal{T}$ to fix the phase of the wavefunctions to be either $+1$ or $-1$ by enforcing:
\begin{equation}
    C_{2z}\mathcal{T} u_{\vec{k},n,\eta}=u_{\vec{k},n,\eta}
\end{equation}
We then fix the relative sign of wavefunctions in opposite flat bands but the same valley with the chiral symmetry operator $C$ as:
\begin{equation}
    C u_{\vec{k},\pm,\eta}=i\eta \pm u_{\vec{k},\mp,\eta}/|\bra{u^*_{\vec{k},\mp,\eta}}C \ket{u_{\vec{k},\pm,\eta}}|
\end{equation}
We fix the relative sign of wavefunctions in opposite bands and opposite valleys with $PH C_{2z}$, where PH a unitary particle hole symmetry operator with:
\begin{equation}
    PH C_{2z} u_{\vec{k},\pm,\eta}=\pm\eta  u_{\vec{k},\mp,-\eta}
\end{equation}
Finally, we use time-reversal symmetry to fix the relative sign between wavefunctions at opposite $\vec{k}$, in opposite valleys, but within the same band:
\begin{equation}
    \mathcal{T} u_{\vec{k},n,\eta}=u_{-\vec{k},n,-\eta}
\end{equation}

\section{Gap Equation at $T=0$}\label{sec:GapT0}
In this appendix we will discuss the self consistency equations we solve to obtain our $T=0$ solutions. In general, we write the Hamiltonian in a Nambu basis as:
\begin{equation}
\mathcal{H}
_{\vec{k}}=    \begin{pmatrix}
        c^\dagger_{\vec{k},+}& c_{-\vec{k},-}
    \end{pmatrix}
    \begin{pmatrix}
        \xi_{\vec{k},+} & \Delta(\vec{k})\\ \Delta(\vec{k})^\dagger & -\xi_{-\vec{k},-}
    \end{pmatrix}\begin{pmatrix}
        c_{\vec{k},+}\\ c^\dagger_{-\vec{k},-}
    \end{pmatrix}
\end{equation}
Where we have suppressed spin and band indices, and both $\xi_{\vec{k},\pm}$ and $\Delta_{\vec{k}}$ are matrices in band and spin space. $\xi_{\vec{k},\pm}$ represents the normal state dispersion in the $\pm$ valleys, which we take to be spin polarized and renormalized by Coulomb interactions as described in App.~\ref{sec:NormalState}. $\Delta_{\vec{k}}$ can be expressed as:
\begin{equation}
    \Delta_{\vec{k}}^{\alpha,\eta;\beta,-\eta}= \frac{1}{N}\sum_{\vec{k},\vec{k'}}\chi_{\vec{k},\vec{k'}}\lambda_{\vec{k},\vec{k'}}^{\alpha,\eta;\gamma,\eta'}\left(\langle c_{-\vec{k}}c_{\vec{k}}\rangle^T\right)^{\gamma,\eta';\delta,-\eta'}\left(\lambda_{-\vec{k},\vec{k'}}^T\right)^{\delta,-\eta';\beta,-\eta}
\end{equation}
In the above, $\lambda_{\vec{k},\vec{k'}}^{\alpha,\eta;\gamma,\eta'}$ represent form factors of some matrix elements which could represent either phonons or fluctuations projected into the flat bands and may be valley diagonal or off diagonal. $V_{\vec{k},\vec{k'}}$ is an isotropic potential which we will generally take to be attractive and flat for phonons and attractive with some lorentzian form for fluctuation mediated pairing.
Since we will be assuming interactions with strength less than the scale of the coulomb interactions, we will treat the polarized spin flavor which is fully occupied at $\nu=2$ as a spectator and assume the pairing is zero in these bands. The self consistency condition we solve at $T=0$ is:
\begin{equation}
    \langle c_{-\vec{k},\alpha,-}c_{\vec{k},\beta,+}\rangle=U_{\vec{k}}^*\chi_{\vec{k}}U_{\vec{k}}^T
\end{equation}
Where $U_{\vec{k}}$ is defined as the unitary operator such that:
\begin{equation}
    U^\dagger_{\vec{k}}\mathcal{H}_{\vec{k}}U_{\vec{k}}=D_{\vec{k}}
\end{equation}
Here, $D_{\vec{k}}$ is a diagonal matrix with the Fermi-Dirac functions of eigenvalues of $\mathcal{H}_{\vec{k}}$ at $T=0$ as its diagonal entries. $\chi_{\vec{k}}$ is the matrix with Fermi-Dirac functions at $T=0$ K of the entries of $D$ on the diagonal. We also must impose Fermi-Dirac statistics as a constraint on our solutions. We enforce this constraint at each iteration by splitting $\langle c_{-\vec{k},\alpha,-}c_{\vec{k},\beta,+}\rangle$ into components which  go as either $\eta_x$ in valley space (denoted $E_{\vec{k}}$) or $\eta_y$ in valley space (denoted as $O_{\vec{k}}$) depending on whether the pairing is even or odd under $\vec{k}\rightarrow-\vec{k}$ and the antisymmetry or symmetry of the band indices as:
\begin{equation}
    O_{\vec{k}}=\frac{1}{2}\left(\langle c_{-\vec{k},\alpha,-}c_{\vec{k},\beta,+}\rangle+\langle c_{\vec{k},\beta,-}c_{-\vec{k},\alpha,+}\rangle\right) \qquad E_{\vec{k}}=\frac{1}{2}\left(\langle c_{-\vec{k},\alpha,-}c_{\vec{k},\beta,+}\rangle-\langle c_{\vec{k},\beta,-}c_{-\vec{k},\alpha,+}\rangle\right)
\end{equation}
Our iterative procedure then proceeds as follows. At the zeroth iteration, an ansatz for $\langle c_{-\vec{k},\alpha,-}c_{\vec{k},\beta,+}\rangle$ satisfying the desired symmetries is selected. Then at each iteration, the chemical potential is adjusted to give the desired filling, which we take to be $\nu=2.5$ in our numerics.
$U_{\vec{k}}$ and the resulting functions $O_{\vec{k}}$ and $E_{\vec{k}}$ are then computed and plugged back into $\Delta_{\vec{k}}$, (which also is guaranteed to obey Fermi-Dirac statistics assuming our generalized form factors obey time reversal symmetry). $\Delta_{\vec{k}}$ is then used to compute the new $U_{\vec{k}}$, and the procedure is repeated until convergence is reached in $\Delta_{\vec{k}}$ and $\mu$. In practice, in our $T=0$ numerics, we take $\vec{q}=\vec{k}-\vec{k'}$ to only be summed over the first Brillouin zone when we consider fluctuation mediated superconductivity, an assumption justified for our fluctuation mediated SC by $\chi(\vec{q})$ falling off as $\frac{1}{|\vec{q}|^2}$ near the first Brillouin zone edge. For phonon mediated superconductivity, we include an additional shell of the 6 nearest Brillouin zones in our sum over $\vec{q}$. Including more shells may reduce the needed coupling, though we expect the leading instability of $A_1$ phonons should be unchanged.

\section{Linearized Gap Equation at $T_c$}\label{SuperconductingGapEquation}
In this appendix, we will describe how we compute solutions to the linearized gap equation at $T_c$. As in App.~\ref{sec:GapT0}, we will assume a spin polarized normal state and only consider superconducting instabilities within a single spin flavor. We recall that for the case of fluctuation-mediated superconductivity, we couple electrons to bosonic modes ($j=1,2,\dots$) as, e.g., in Eq.~(4), with $\lambda^j_{\alpha,\eta;\alpha',\eta'}$ capturing the symmetries broken by the corresponding order parameter. In order to compactly write down the linearized gap equation, it is convenient to express $\lambda^j$ as
\begin{equation}
    \left(\lambda^j_{\alpha,\eta;\alpha',\eta'}\right)_{\vec{k},\vec{k}'}=A^{\alpha,\eta;\alpha',-\eta}_{\vec{k},\vec{k}'}\delta_{\eta,-\eta'}+B^{\alpha,\eta;\alpha',\eta}_{\vec{k},\vec{k}'}\delta_{\eta,\eta'}. \label{RewritingofMatrixElements}
\end{equation}
Here we also include the momentum dependence of the matrix elements, which arises when we study phonons and order parameter fluctuations projected from the sublattice basis to the band basis. In \equref{RewritingofMatrixElements}, $A_{\vec{k},\vec{k}'}$ are the valley off diagonal pieces of the form factor $\lambda^j_{\alpha,\eta;\alpha',\eta'}$ and $B_{\vec{k},\vec{k}'}$ is the valley diagonal pieces.
With this notation in hand, the linearized gap equation we solve is
\begin{equation}\label{eq:lingap}
    \left(\Delta(\vec{k})^\dagger\right)^{\alpha,-;\beta+}=\sum_{\vec{q}}\chi_{\vec{q}}\left(\mathcal{G}_{\vec{k}-\vec{q}}^{\delta-;\gamma+}B_{\vec{k}-\vec{q},\vec{q}}^{\gamma+;\beta+}(B_{-\vec{k}+\vec{q},-\vec{q}}^T)^{\alpha-;\delta-}-\mathcal{G}_{-\vec{k}+\vec{q}}^{\gamma-;\delta+}A_{\vec{k}-\vec{q},\vec{q}}^{\gamma-;\beta+}(A_{-\vec{k}+\vec{q},-\vec{q}}^T)^{\alpha-;\delta+}\right), 
\end{equation}
where the Greens function $\mathcal{G}^{\alpha+;\beta-}_{\vec{k}-\vec{q}}$ defined by 
\begin{equation}
\begin{split}
    \mathcal{G}^{\alpha+;\beta-}_{\vec{k}}=\frac{1}{2A_{m}}&\left(\frac{\Delta_{\vec{k}}^{00}}{2|\xi_{\vec{k},0}|}\left(1-2n_F(|\xi_{\vec{k},0}|)\right)(\sigma_0+\sigma_z)^{\alpha\beta}+ \frac{\Delta_{\vec{k}}^{11}}{2|\xi_{\vec{k},1}|}\left(1-2n_F(|\xi_{\vec{k},1}|)\right)(\sigma_0-\sigma_z)^{\alpha\beta}+\right.\\&\left.\frac{\Delta_{\vec{k}}^{01}}{\xi_{\vec{k},0}+\xi_{\vec{k},1}}(n_F(-\xi_{\vec{k},1})-n_F(\xi_{\vec{k},0}))(\sigma_x+i\sigma_y)^{\alpha\beta}+\frac{\Delta_{\vec{k}}^{10}}{\xi_{\vec{k},0}+\xi_{\vec{k},1}}(n_F(-\xi_{\vec{k},1})-n_F(\xi_{\vec{k},0}))(\sigma_x-i\sigma_y)^{\alpha\beta}\right).
\end{split}
\end{equation}
Here $\Delta^{\alpha\beta}_{\vec{k}}$ denote the pairing in band space where $\alpha,\beta=0,1$ label the upper and lower flat band.
Finding a solution to the above equation then amounts to computing the right-hand side of Eq.~(\ref{eq:lingap}), diagonalizing it in the space of momenta, Nambu index, and band index, and looking at the eigenvectors which attain eigenvalue 1 for some value of $T$. To enforce Fermi-Dirac statistics, we solve the above equation on half of the moir\'e Brillouin zone. We also exclude the edge points in our linearized gap equation computations for phonons and projected order fluctuations. We expect including these points would reduce the needed coupling to obtain a finite $T_c$ (or reduce $T_c$ for fixed coupling) but not change the leading instabilities. 

\section{Additional statements about superconductivity and phonons}

\begin{table}[tb]
\begin{center}
\caption{Generalization of Table II of the main text, where we also indicate the dominant superconducting orders ($\bar{\Delta}$) in the microscopic basis, obtained by applying Eq.~(7) in the sublattice basis. The phonon modes refer to the sublattice-basis form $\bar{\lambda}_j$ of the coupling, cf.~Eq.~(8), and ``g-nematic'' stands for the (intravalley) graphene nematic state of \refcite{OurNematicTheory}, which has the same coupling as the $E_2$ phonon.}
\label{ExtendedPreferredPairingStates}
\begin{ruledtabular}
 \begin{tabular} {ccccccc} 
\multicolumn{3}{c}{Fluctuating Order} & \multicolumn{2}{c}{Leading SC (band)} & \multicolumn{2}{c}{Leading SC (microscopic)} \\
\cline{1-3} \cline{4-5} \cline{6-7}
type & $\lambda^j$ & $\bar{\lambda}^j$ & $\Delta_{\vec{k},\eta}$ & IR & $\bar{\Delta}_{\vec{k},\eta}$ & IR
\\ \hline
T-IVC/$A_1,B_1$ phonon & $\sigma_0 \eta_{x,y}$ & $\rho_x \eta_{x,y}$ & $\sigma_y \delta_{\vec{k}}$ & $A_2$ & $\rho_z \eta  \delta_{\vec{k}}$ & $A_2$ \\
K-IVC & $\sigma_y \eta_{x,y}$ & $\rho_y \eta_{x,y}$ & $\sigma_0 \eta \delta_{\vec{k}}$ & $B_1$ & $ \rho_0 \eta \delta_{\vec{k}} $ & $B_1$ \\
SLP$+$ & $\sigma_y \eta_{z}$ & $\rho_z \eta_0$ & $\sigma_y \delta_{\vec{k}}$, $\sigma_0 \eta \delta_{\vec{k}}$ & $A_2$, $B_1$ & $\rho_z\eta \delta_{\vec{k}}$, $\rho_0\eta \delta_{\vec{k}}$  & $A_2$, $B_1$ \\
SLP$-$ & $\sigma_y \eta_{0}$ & $\rho_z\eta_z$ & $\sigma_x \eta \delta_{\vec{k}}$, $\sigma_z \eta \delta_{\vec{k}}$ & $B_2$, $B_1$ & $(\rho_y,\rho_x \eta)$ & $E_1$ \\
N-IVC & --- & $\eta_{x,y}\rho_{0,z}$ & --- & --- & $(\rho_y,\rho_x \eta)$ & $E_1$ \\
g-nematic/$E_2$ phonon & --- & $(\eta_z\rho_y, -\rho_x)$ & --- & --- & $ \rho_0 \eta \delta_{\vec{k}} $ & $B_1$
 \end{tabular}
 \end{ruledtabular}
\end{center}
\end{table}

\subsection{Generalization to sublattice basis}\label{ApplyInSublatticeBasis}
Due to the basis independent form of the (anti)commutator relation in Eq.~(7), it can be readily applied in any basis. As we also study in the numerics of the main text momentum-independent coupling matrices $\bar{\lambda}_j$ in the microscopic sublattice basis, it seems natural to also apply the commutator relation in that basis. Upon noting that the additional projection onto the flat bands does, in general, not commute with the order parameters, it is clear that applying Eq.~(7) can only provide approximate guidance even in the strict flat-band limit. Notwithstanding these approximations, the results, summarized in \tableref{ExtendedPreferredPairingStates}, agree well with the numerics shown in Fig.~4 of the main text. In the case of N-IVC fluctuations, the listed $E_1$ superconductor is the option where the highest number of components obey Eq.~(7), while all components obey it in all other cases.

\subsection{Electron phonon coupling in TTG}\label{ElPhCouplTTG}
As it exhibits three layers, the discussion of the layer structure of the phonon modes in TTG requires additional comments. 
Starting from uncoupled optical $A_1$, $B_1$, and $E_2$ phonons in the three layers of TTG, we can decompose each of these modes into two mirror-even ($\mu=e_1,e_2$) and one mirror-odd ($\mu=o$) contributions,
\begin{equation}
    \vec{v}_{e_1} = \frac{1}{\sqrt{3}}\begin{pmatrix} 1 \\ 1 \\ 1 \end{pmatrix}, \quad \vec{v}_{e_2} = \frac{1}{\sqrt{6}} \begin{pmatrix} 1 \\ -2 \\ 1 \end{pmatrix}, \quad \vec{v}_{0} = \frac{1}{\sqrt{2}} \begin{pmatrix} 1 \\ 0 \\ -1 \end{pmatrix}.
\end{equation}
Upon projection into the mirror-even electronic sectors, forming the relevant low-energy flat-band degrees of freedom, the mode $\vec{v}_{0}$ vanishes completely (due to mirror-symmetry), while the first two survive. Their respective projected coupling is of the form of Eq.~(8) with $\vec{v}_{e_1} = (1,1)^T/\sqrt{3}$ and $\vec{v}_{e_2} = (1,-2)^T/\sqrt{6}$. 

\subsection{Electron-phonon matrix elements}\label{ElPhMatrixElements}
In this appendix, we analyze the momentum-independent terms of the electron-phonon coupling matrices $\lambda_{\vec{k},\alpha,\eta;\vec{k}',\alpha'\eta'}^{g,j,\mu}$ in Eq.~(9). As a result of $C_{3z}$ symmetry, the coupling terms of $E_{2}$ cannot have a momentum-independent component and so we focus on $g=A_1, B_1$. Let us expand in Pauli matrices in band and valley space,
\begin{equation}
    \lambda_{\vec{k},\alpha,\eta;\vec{k}',\alpha'\eta'}^{g,\mu} = \sum_{j_1,j_2} c^{g,\mu}_{j_1,j_2} (\sigma_{j_1})_{\alpha,\alpha'}  (\eta_{j_2})_{\eta,\eta'} + \mathcal{O}(\vec{k},\vec{k}'),  \label{ProjectedFormFactors}
\end{equation}
where Hermiticity implies $c^{g,\mu}_{j_1,j_2} \in \mathbbm{R}$.
The combination of $U(1)_v$ (valley-charge conservation), $C_{2z}$, and $\Theta$ implies that only $c^{A_1,\mu}_{j_1,x}$, $c^{B_1,\mu}_{j_1,y}$, $j_1=0,x,z$ can be non-zero. Chiral symmetry $C$ has the representation $\rho_z$ and $\eta_z\sigma_y$ in the sublattice and band basis, respectively. As $\rho_z$ anti-commutes with both $\Lambda_{A_1} = \eta_x \rho_x$ and $\Lambda_{B_1} = \eta_y \rho_x$, their band projections in \equref{ProjectedFormFactors} also have to anti-commute with $\eta_z\sigma_y$; this leaves us with $c^{A_1,\mu}_{0,x}$ and $c^{B_1,\mu}_{0,y}$ as the only non-zero terms. Furthermore, the unitary particle-hole symmetry $P$ anti-commutes with the layer-even ($\mu=+$) and commutes with the layer-odd ($\mu=-$) modes. Being represented by $-i \eta_z\sigma_y$, this is inconsistent with $c^{A_1,-}_{j_1,x},c^{B_1,-}_{j_1,y}\neq 0$, which thus have to vanish. This is in line with our numerics, where we find very small projections of the layer-odd $A_1$ and $B_1$ modes. Their layer-even counterparts, however, are consistent with $P$ if only $c^{A_1,\mu}_{0,x}$ and $c^{B_1,\mu}_{0,y}$ are non-zero. Taken together, we find
\begin{equation}
    \lambda_{\vec{k},\alpha,\eta;\vec{k}',\alpha'\eta'}^{g,-} =  \mathcal{O}(\vec{k},\vec{k}'),\, g=A_1,B_1, \quad \lambda_{\vec{k},\alpha,\eta;\vec{k}',\alpha'\eta'}^{A_1,+} = \sigma_0 \eta_x + \mathcal{O}(\vec{k},\vec{k}'), \quad \lambda_{\vec{k},\alpha,\eta;\vec{k}',\alpha'\eta'}^{B_1,+} = \sigma_0 \eta_y + \mathcal{O}(\vec{k},\vec{k}').
\end{equation}

\section{Pairing for other normal-state orders}\label{sec:Nopolarization}

In the main text, we have discussed pairing in the case of a spin polarized or spin-valley locked normal state. We here comment on the consequences for superconductivity for two other, plausible normal-state scenarios.

\subsection{T-IVC \& SP order}
Given the current insights from experiment, the most natural alternative scenario is that the normal state exhibits both T-IVC \cite{https://doi.org/10.48550/arxiv.2303.00024} and spin polarization \cite{Lake_2022,morissetteElectronSpinResonance2022} simultaneously. The projection to the remaining two active flavor degrees of freedom is given by 
\begin{equation}
    P_{\nu=2}=\frac{1}{4}\left(1+s_z\right)\left(1+\eta_x\rho_x\right). \label{Projector1}
\end{equation}
Increasing $\nu$ beyond $\nu=2$ will lead to a metallic state with two non-degenerate bands $\alpha=\pm$ coming from the original flat-band manifold. Let us denote the associated creation operators by $c^\dagger_{\vec{k},\alpha}$, which have one index less than the associated operators discussed in the main text since valley is not a good quantum number anymore. The superconducting order parameter is a $2\times 2$ matrix, coupling to the electrons as $\sum_{\vec{k},\alpha,\alpha'} c^\dagger_{\vec{k},\alpha} \left( \Delta_{\vec{k}}\right)_{\alpha,\alpha'} c^\dagger_{-\vec{k},\alpha'} + \text{H.c.}$, and thus has to obey $\Delta_{\vec{k}} = -\Delta_{-\vec{k}}^T$. As the projector in \equref{Projector1} commutes with $C_{2z}$ (in fact, also with $C_{2x}$ and $C_{3z}$), all pairing states must still be either even or odd under $C_{2z}$ (transform under one of the IRs of $D_{6}$ or $C_6$). Since \equref{Projector1} projects onto the subspace where $\eta_x\rho_x$ is $-1$, it holds $C_{2z}$: $c^\dagger_{\vec{k},\alpha} \rightarrow -c^\dagger_{-\vec{k},\alpha}$ and, hence,
\begin{equation}
    C_{2z}: \quad \Delta_{\vec{k}} \quad \longrightarrow \quad \Delta_{-\vec{k}} = - \Delta^T_{\vec{k}},
\end{equation}
which is the analogue of Eq.~(2) of the main text. As before, all $C_{2z}$-even states must be entirely band-off-diagonal, $\Delta_{\vec{k}} = \delta_{\vec{k}} \sigma_y$. However, since the number of active degrees of freedom is reduced, there are more restrictions: all $C_{2z}$-odd superconductors must have zeros in the Brillouin zone due to $\Delta_{\vec{k}} = -\Delta_{-\vec{k}}^T = -\Delta_{-\vec{k}}$.

For completeness and to conveniently address energetics, we extend the discussion to the microscopic sublattice basis. Let $\bar{\Delta}_{\vec{k}}$ be the corresponding superconducting order parameter---an $8\times8$ matrix in sublattice, valley, and spin space. Then pairings are constrained to obey
\begin{equation}\label{CondProj}
    P_{\nu=2} \bar{\Delta}_{\vec{k}} s_y\eta_x  P_{\nu=2}^T = \bar{\Delta}_{\vec{k}}s_y\eta_x.
\end{equation}
The order parameters which are compatible with \equref{CondProj} will all be spin triplets. 
The $C_{2z}$-even states, i.e., order parameters transforming under $A_2$, $E_2$, or $A_1$, will have the form (suppressing $\vec{k}$-dependencies) $\bar{\Delta}_{\vec{k}}\sim P_{\nu=2}s_x\eta_z\rho_z$; in line with our symmetry arguments above, one can check that they will go as $\sigma_y$ in band space and thus be purely band off diagonal in the subspace defined by $P_{\nu=2}$. The pairings which are odd under $C_{2z}$ include the $B_1$ and $B_2$ pairings with $\bar{\Delta}_{\vec{k}}\sim P_{\nu=2}s_x\rho_0$, and $E_1$ pairings with $\bar{\Delta}_{\vec{k}}\sim P_{\nu=2}s_x(\rho_x,\rho_y\eta_z)$ previously discussed in our main text; however, as pointed out above and unlike in the main text, the $C_{2z}$-odd pairings in both the band basis and sublattice basis are no longer allowed to have a component without a sign change since only the momentum odd components of the $B_1$, $B_2$, and $E_2$ pairings survive projection $P_{\nu=2}$. 

Since only the band-off-diagonal $A_2$ state can have a non-sign-changing order parameter, a superconducting state satisfying the criterion around Eq.~(7) of the main text can only be this state (or none). We have studied which of the pairing mechanisms survive the projection and whether they favor or disfavor $A_2$ pairing, see Table~\ref{TableTIVCSCSP}.
We find that $A_1$ phonons, T-IVC fluctuations, and spin fluctuations all provide an attractive pairing potential, and if any of these have large enough couplings to overcome the normal state band splitting, the $A_2$ triplet pairing is the leading instability, as in the main text. Furthermore, due to the fact that the remaining bands after reconstruction, as described by the projector $P_{\nu=2}$, are not degenerate (there is no remaining spin symmetry to guarantee degeneracy), a Bogoliubov Fermi surface or a fully gapped state and, thus, a transition from nodal to gapped as a function of filling are possible depending on parameters (similar to our discussion in the main text).

\begin{table}[H]
    \centering
    \begin{tabular}{c|c|c|c|c|c|c}
       $\Delta_{\vec{k}}$ & IR of $D_6$ &\begin{tabular}{@{}c@{}}T-IVC/$A_1$ phonon\\ $\eta_x\rho_x$ \end{tabular}  
       
    & \begin{tabular}{@{}c@{}}quantum spin Hall\\ $s_z\eta_z\rho_z$\end{tabular}     &   \begin{tabular}{@{}c@{}}spin polarized\\ $s_z$ \end{tabular} & \begin{tabular}{@{}c@{}}N-IVC\\ $\eta_x(\rho_0,\eta_z\rho_z)$ \end{tabular}  & \begin{tabular}{@{}c@{}}quantum Hall\\ $\eta_z\rho_z$ \end{tabular} \\
       \hline
       $P_{\nu=2}\left(s_x\eta_z\rho_z\delta_{\vec{k}}\right)$ & $A_2$ & \cmark & \xmark & \cmark &\xmark &\xmark
    \end{tabular}
    \caption{We list the possible pairing glues which are compatible with a T-IVC+SP normal state (i.e., the interactions survive projection to the space of the upper T-IVC bands of a single spin flavor). We denote interactions which will generate an attractive interaction for the $A_2$ pairing with a \cmark~and interactions which will generate a repulsive interaction with a \xmark.}
    \label{TableTIVCSCSP}
\end{table}

\subsection{T-IVC normal state}
We will now consider a simpler normal state which leaves twice the number of degrees of freedom as the previous normal state we considered. In particular, we can consider a strong coupling T-IVC normal state with projector of the form:
\begin{equation}
    P_{\nu=2}=\frac{1}{2}\left(1+\eta_x\rho_x\right)
\end{equation}

In contrast to the case for a normal state with coexisting T-IVC and spin-polarized order, there are now more possible pairing options and singlet pairing is once again possible. We can classify the possibilities as pairings which are triplet, singlet, and by IRs of the point group. We find the possible pairings include triplet $A_1$ and $A_2$ pairings and singlet $B_1$ and $B_2$ pairings with:
\begin{equation}
    \Delta_{\vec{k}}\sim P_{\nu=2}s_x\eta_z\rho_z \quad \Delta_{\vec{k}}\sim P_{\nu=2}s_0\eta_z\rho_z
\end{equation}
triplet $B_1$ and $B_2$ pairings and singlet versions of our $A_1$ and $A_2$ states with:
\begin{equation}
    \Delta_{\vec{k}}\sim P_{\nu=2}s_x\eta_0\rho_0 \quad \Delta_{\vec{k}}\sim P_{\nu=2}s_0\eta_0\rho_0
\end{equation}
and triplet $E_1$ pairing and singlet $E_2$ pairing with:
\begin{equation}
    \Delta_{\vec{k}}\sim P_{\nu=2}s_x(\eta_0\rho_x,\eta_z\rho_y) \quad \Delta_{\vec{k}}\sim P_{\nu=2}s_0(\eta_0\rho_x,\eta_z\rho_y)
\end{equation}

Of the above, the only options which are not enforced to have a sign change are our purely inter-band $A_2$ triplet pairing, the $A_1$ singlet pairing, and the $E_2$ singlet pairing. Since these pairings do not have a sign change, they are the only possible candidates for the criterion around Eq.~(7) of the main text and we have enumerated the possible pairing glues for these s-wave states in Table~\ref{TableTIVCSC}.
\begin{table}[H]
    \centering
    \begin{tabular}{c|c|c|c|c|c|c}
       $\Delta_{\vec{k}}$ & IR of $D_6$  &\begin{tabular}{@{}c@{}}T-IVC/$A_1$ phonon\\ $\eta_x\rho_x$ \end{tabular}  
       
    & \begin{tabular}{@{}c@{}}quantum spin Hall\\ $\vec{s}\eta_z\rho_z$\end{tabular}     &   \begin{tabular}{@{}c@{}}Spin polarized\\ $\vec{s}$ \end{tabular} & \begin{tabular}{@{}c@{}}N-IVC\\ $\eta_x(\rho_0,\eta_z\rho_z)$ \end{tabular}  & \begin{tabular}{@{}c@{}}quantum Hall\\ $\eta_z\rho_z$ \end{tabular} \\
       \hline
       $P_{\nu=2}\left(s_x\eta_z\rho_z\delta_{\vec{k}}\right)$ & $A_2$ (triplet) & \cmark & \xmark & \cmark &\xmark &\xmark\\
       \hline
       $P_{\nu=2}\left(s_0\eta_0\rho_0\delta_{\vec{k}}\right)$& $A_1$ (singlet)  & \cmark & \cmark & \xmark &\cmark &\xmark\\
       \hline
       $P_{\nu=2}\left(s_0(\eta_0\rho_x,\eta_z\rho_y)\right)$& $E_2$ (singlet)  & \cmark & \xmark & \xmark &\cmark &\cmark\\
    \end{tabular}
    \caption{We list the possible pairing glues which are compatible with a T-IVC normal state (ie the interactions survive projection to the space of the upper spin-degenerate T-IVC bands). We denote interactions which will generate an attractive interaction with a \cmark and interactions which will generate a repulsive interaction with a \xmark.}
    \label{TableTIVCSC}
\end{table}

We find in this case that all of the pairing glues which are attractive for our $A_2$ triplet pairing are also attractive for one of the singlet pairings, except for spin fluctuations. Therefore, we can say that if the pairing is triplet for a spin-degenerate T-IVC normal state, the leading instability is likely to be our $A_2$ pairing provided the pairing glue interaction is sufficiently strong and spin fluctuations may play an important role in energetically favoring this state. In this case, we expect that the phenomenology of Bogoliubov Fermi surfaces and a nodal to gapped transition as a function of interaction strength will again apply.

\section{More Superconducting Instabilities}\label{sec:MoreInstabilities}
\begin{figure}[tb]
    \centering
    \includegraphics[width=\linewidth]{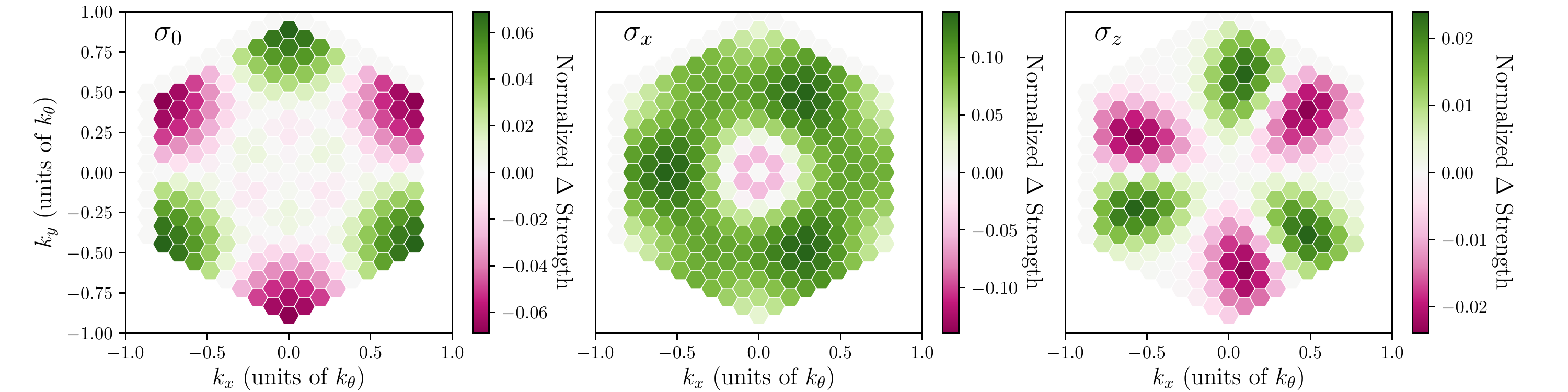}
    \caption{Highest eigenvalue pairing obtained from linearized gap equation at $T=16$ K, for SLP- and T-IVC fluctuations. The pairing transforms under the $B_2$ representation of the point group.}
    \label{fig:B2Pairing}
\end{figure}

In this appendix, we will discuss the superconducting instabilities we find beyond the $A_2$ and $B_1$ states shown in Figs. 2 and 3 of the main text and focus on the other leading instabilities we find in the presence of fluctuations of different particle hole orders. For SLP$-$ fluctuations, we find the $B_2$ state can be favored over the $B_1$ when the strength of T-IVC fluctuations are on the same order as SLP$-$ fluctuations, as shown in Fig. 4. We show the $B_2$ state for parameter value $\theta_{fluc.}\approx\frac{\pi}{4}$ in Fig.~\ref{fig:B2Pairing}.
For N-IVC fluctuations as well as for SLP$-$ fluctuations, we find the $E_2$ is the leading instability, as shown in Fig. 6. We show the two components of the $E_2$ state for parameter value $\theta_{fluc.}\approx\frac{\pi}{2}$ in Fig. 4.
\begin{figure}[tb]
    \centering
    \includegraphics[width=\linewidth]{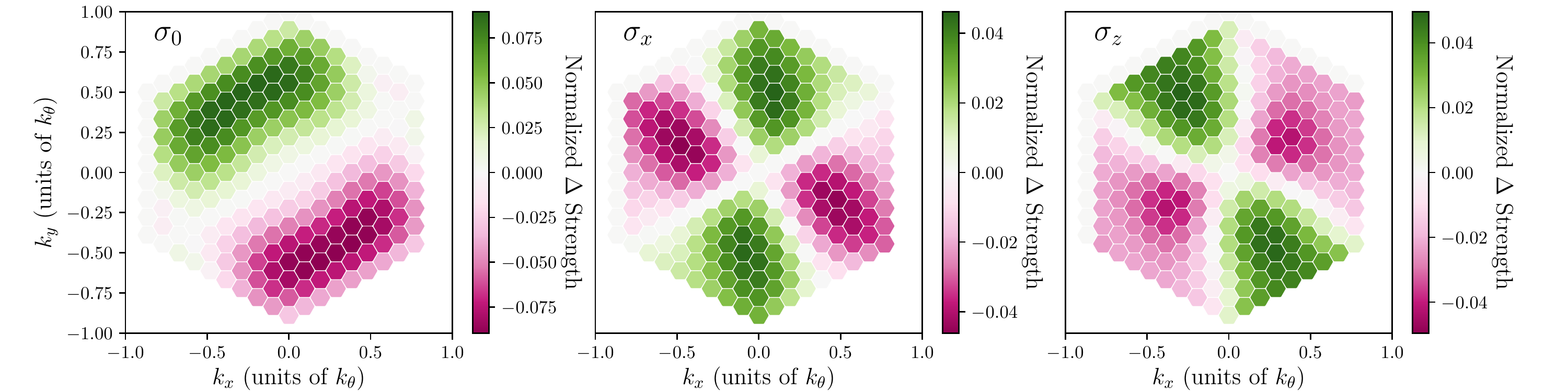}
    \caption{First component of the highest eigenvalue pairing obtained from linearized gap equation at $T=16$ K for just N-IVC fluctuations. The pairing transforms under the $E_1$ representation of the point group. The component shown here is degenerate with the other basis functions which transform under $C_3$ symmetry shown in Fig.~\ref{fig:E1bPairing}.}
    \label{fig:E1aPairing}
\end{figure}
\begin{figure}[H]
    \centering
    \includegraphics[width=\linewidth]{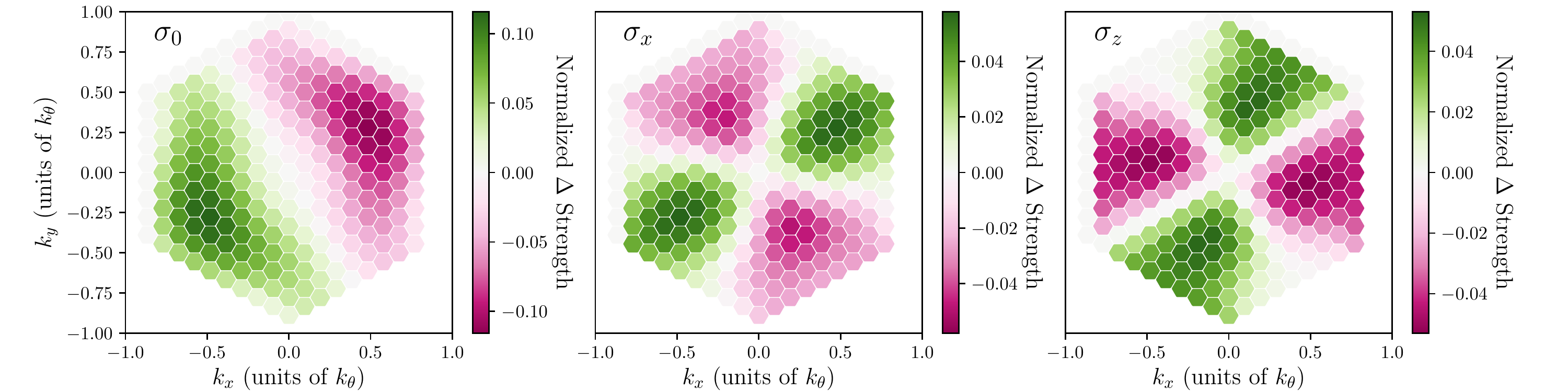}
    \caption{Second component of the highest eigenvalue pairing obtained from linearized gap equation at $T=16$ K for just N-IVC fluctuations. The pairing transforms under the $E_1$ representation of the point group. The component shown here is degenerate with the other basis functions which transform under $C_3$ symmetry shown in Fig.~\ref{fig:E1aPairing}.}
    \label{fig:E1bPairing}
\end{figure}
We point out that each component of the $E_1$ pairing shown in Figs.~\ref{fig:E1aPairing} and \ref{fig:E1bPairing} may by themselves be nodal, assuming the pieces of each pairing which are proportional to $\sigma_x$ in band space are smaller than the band splitting. In general, we expect the lowest energy pairing at $T=0$ will be the chiral $E_1$ state which would be fully gapped; however, in the presence of sufficient strain, a single basis function of the $E_1$ pairing can be favored over the chiral state, offering another route to nodal superconductivity in the presence of N-IVC fluctuations. 
\end{appendix}
\end{document}